\documentclass[%
reprint,superscriptaddress,
 amsmath, amssymb, aps, pre,
floatfix]{revtex4-2}
\UseRawInputEncoding
\usepackage{graphicx}
\usepackage{dcolumn}
\usepackage{bm}
\usepackage{hyperref}
\usepackage{xcolor}
\usepackage[caption = false]{subfig}
\usepackage[utf8]{}
\usepackage{float}
\usepackage{amsmath}
\usepackage{amsfonts}
\begin{document}

\title{Computing macroscopic reaction rates in reaction-diffusion systems \\ 
	using Monte Carlo simulations}

\author{Mohamed Swailem} \email{mswailem@vt.edu}
\affiliation{Department of Physics \& Center for Soft Matter and Biological Physics, MC 0435, Robeson Hall, 
		850 West Campus Drive, Virginia Tech, Blacksburg, VA 24061, USA}
\author{Uwe C. T{\"a}uber} \email{tauber@vt.edu}
\affiliation{Department of Physics \& Center for Soft Matter and Biological Physics, MC 0435, Robeson Hall, 
		850 West Campus Drive, Virginia Tech, Blacksburg, VA 24061, USA}
\affiliation{Faculty of Health Sciences, Virginia Tech, Blacksburg, VA 24061, USA}

\date{\today}

\begin{abstract}
Stochastic reaction-diffusion models are employed to represent many complex physical, biological, 
societal, and ecological systems. 
The macroscopic reaction rates describing the large-scale, long-time kinetics in such systems are effective,
scale-dependent renormalized parameters that need to be either measured experimentally or computed by
means of a microscopic model. 
In a Monte Carlo simulation of stochastic reaction-diffusion systems, microscopic probabilities for specific
events to happen serve as the input control parameters. 
To match the results of any computer simulation to observations or experiments carried out on the 
macroscale, a mapping is required between the microscopic probabilities that define the Monte Carlo 
algorithm and the macroscopic reaction rates that are experimentally measured. 
Finding the functional dependence of emergent macroscopic rates on the microscopic probabilities 
(subject to specific rules of interaction) is a very difficult problem, and there is currently no systematic,
accurate analytical way to achieve this goal. 
Therefore, we introduce a straightforward numerical method of using lattice Monte Carlo simulations to 
evaluate the macroscopic reaction rates by directly obtaining the count statistics of how many events 
occur per simulation time step. 
Our technique is first tested on well-understood fundamental examples, namely restricted birth processes,
diffusion-limited two-particle coagulation, and two-species pair annihilation kinetics. 
Next we utilize the thus gained experience to investigate how the microscopic algorithmic probabilities 
become coarse-grained into effective macroscopic rates in more complex model systems such as the 
Lotka--Volterra model for predator-prey competition and coexistence, as well as the rock-paper-scissors
or cyclic Lotka--Volterra model as well as its May--Leonard variant that capture population dynamics with
cyclic dominance motifs. 
Thereby we achieve a more thorough and deeper understanding of coarse-graining in spatially extended
stochastic reaction-diffusion systems and the nontrivial relationships between the associated microscopic
and macroscopic model parameters, with a focus on ecological systems.
The proposed technique should generally provide a useful means to better fit Monte Carlo simulation 
results to experimental or observational data.
\end{abstract}

\maketitle

\section{\label{section:Intro} Introduction}

Reeaction-diffusion systems, which describe the dynamical evolution of different species through local 
reactions that alter their identity, and that propagate diffusively (in the continuum) or through hopping
(among discrete lattice sites), arise in many descriptive models of (bio-)chemical, nuclear, and particle 
reactions,  population evolution in ecological systems, spreading of contagious diseases in epidemiology, 
the physics and materials science of pattern formation, and evolutionary game theory in mathematics, 
computer science, economics, and sociology \cite{intro1, intro2, intro3, intro4, intro5, intro6, intro7, 
intro8, intro9, intro10, intro11, intro12, intro13, intro14, intro15, intro16, uwesbook}. 
However, complete exact analytical solutions are seldom obtainable for theoretical formulations of such 
systems due to the presence of many degrees of freedom, nonlinear interactions, and the underlying
stochasticity. 
Thus a statistical approach is often more suitable for investigating these systems quantitatively. 
Reactive particle models are frequently simplified rather drastically by employing a ``mass-action'' 
factorization approximation which ignores the effects of stochastic fluctuations such as demographic 
noise and the presence of spatio-temporal correlations \cite{intro13, intro14, intro16, mf1, mf2, mf3, 
mf4, lotka, volterra, mf5, mf6, mf7}. 
Consequently, these mean-field type approximations fail for systems that exhibit strong correlations 
between its constituents as can in fact be induced by the reaction processes themselves, and may entail
the emergence of spatial patterns, coherent oscillatory dynamics, etc. 
In addition, it is well-established that even in situations where a reaction-diffusion system can be 
modeled qualitatively by means of coupled mean-field rate equations, often the associated macroscopic 
reaction rates that enter as basic parameters in these deterministic ordinary or partial differential 
equations differ from the corresponding microscopic reaction probabilities per unit time.
Indeed, the macroscopic rates represent effective scale-dependent, ``renormalized'' quantities that may 
become functions of the elapsed time or reactant particle densities \cite{EffectiveRateEquation}.

``Individual-'' or ``agent-based'' Monte Carlo lattice simulations are commonly used as an efficient 
numerical tool to study the properties of reaction-diffusion systems quantitavily \cite{MC1, intro11, 
intro13, intro14, intro15, intro16, LVme}. 
They fundamentally sample the gain-loss balance master equation that defines the stochastic time 
evolution for the configurational probability of the system under consideration.
Their crucial input parameters are microscopic probabilities that ultimately determine the emergent 
macroscopic long-time behavior of the system, namely its stationary properties, if applicable; possible 
large-scale correlated spatial structures; as well as the average relaxation time it takes to reach the 
steady state; or other important time scales such as characteristic oscillation frequencies. 
These simulations are straightforward to implement and as output generate expectation values of 
relevant observables and correlation functions. 
Other pertinent quantities, e.g., correlation lengths, characteristic time scales, scaling exponents, etc. 
may subsequently be extracted from the correlation functions by means of various data analysis 
techniques \cite{intro11, exponents1, exponents2, exponents3, exponents4, exponents5}.
For certain, typically comparatively simple systems, analytical techniques including field-theoretic 
methods may also accomplish the computation of these observables \cite{Doi, Peliti, ft1, ft2, ft3, ft4, ft5, 
ft6, ft7, ft8, ft9, ft10, Cog5,intro12, intro13, ftexponents1, HongPaper, Goldenfeld1, Goldenfeld2, 
Goldenfeld3, EffectiveRateEquation, Lee1, Lee2, Lee3, Lee4, Lee5, Lee6}.

A large set of various models have been studied extensively using Monte Carlo simulations. 
We just mention a small selection here, relevant to our investigations.
In binary coagulation and annihilation models, the strong fluctuation corrections that drastically alter the 
scaling behavior of the particle density relative to the mean-field approximation in the asymptotic 
long-time region were confirmed numerically \cite{EffectiveRateEquation, CogAnn1, CogAnn2, CogAnn3, CogAnn4, CogAnn5, CogAnn6, CogAnn7}. 
In these pair interaction models, ultimately the system can still be effectively described by a modified rate 
equation but with an effective density-dependent reaction rate, i.e., a scale-dependent ``running'' 
coupling that encodes the emergence of strong particle anticorrelations and depletion zones in low 
dimensions $d \leq 2$ \cite{CogDoi, CogPeliti, Lee2, ft2, Cog1, Cog2, Cog3, Cog4, Cog5, CogAnn6, ft10}.
In the literature, when Monte Carlo simulations are utilized to investigate these systems, the effective 
macrosopic reaction rate was inferred indirectly from measurements of the density decays. 
To our knowledge, no attempt has been made to directly obtain the macroscopic reaction rate by
computational means. 
In their detailed experiments on the binary quenching kinetics of laser-induced excitons in quasi 
one-dimensional carbon nanotubes, Allam \textit{et al.} measured the effective time- and 
density-dependent annhiliation rate \cite{CogExperiment}.
Monson and Kopelman \cite{CogAnn2} eperimentally confirmed the considerably slowed-down density 
decay in diffusion-limited two-species annihilation processes in dimensions $d \leq 4$ owing to species 
segregation and confinement of reactions to narrowing reaction zones localized at the interfaces between 
the inert single-species clusters \cite{Lee1, Lee3, Lee4, CogAnn4, CogAnn6, Ann1, Ann2}.

The Lotka--Volterra predator-prey model represents perhaps one of the simplest reaction-diffusion 
systems relevant for ecology that displays unexpectedly rich macro-scale features \cite{lotka, volterra, 
LV1, LV3, intro14, exponents1, LV2, ftexponents1, intro12}. 
This model exhibits both an active-to-absorbing phase transition, namely an extinction threshold for the
predator species, and the emergence of dynamical spatial patterns in the two-species coexistence phase. 
Both these phenomena are probed by calculations of the static and dynamic correlation functions using 
Monte Carlo simulations \cite{exponents1, intro11, exponents3, exponents4, exponents2, exponents5, 
ftexponents1, intro12, intro13}. 
Stochastic fluctuations connected with the activity or evasion-pursuit fronts continually traversing the 
system modify its characteristic parameters such as the population oscillation frequency and damping, the
diffusivity, and the reaction rates and nonlinear couplings. 
Regardless, investigations of the connection between microscopic and macroscopic rates are scarce in the 
literature. 
Lastly, three-species population dynamics models that implement cyclic dominance motifs such as the 
cyclic Lotka--Volterra or ``rock-paper-scissors'' model, and its May--Leonard variant where predation and
reproduction processes are decoupled, were also studied extensively by means of Monte Carlo simulations
\cite{mf7, RPS1, RPS2, RPS3, RPS4, RPS5, RPS6, RPS7, RPS8, ML1, ML2, ML3, ML4, ML5WithRPSLimit, 
ML6, ML7, ML8, cyclic1, cyclic2, cyclic3, cyclic4, cyclic5, cyclic6, cyclic7}.
Crucially, for the cyclic Lotka--Volterra model the total particle number is conserved, which causes its
large-scale features to drastically differ from the similar May--Leonard realization, where this conservation
law does not apply:
In contrast to the rock-paper-scissors variant, the cyclic May--Leonard model displays the spontaneous 
formation of spiral spatial patterns. 

In this paper we focus on utilizing Monte Carlo simulations to investigate the functional dependences of 
the macroscopic reaction rates on microscopic system parameters as well as elapsed time and / or reactant
densities. 
The effective coarse-grained reaction rates can be directly computed by obtaining the count statistics of the 
number of reactions occurring of a specific type at each simulation time step. 
Numerically calculating the macroscopic rates thus generates the desired mapping between the microscopic 
parameters in the Monte Carlo algorithm and the effective large-scale reaction rates. 
These macroscopic rates may then be plugged into the corresponding rate equations, which constitute a
concise and fairly simple representation of a complex stochastic system. 
We shall demonstrate that at least for some models, this technique is consistent with known results in the 
literature, and hence offers a viable means to effectively include subtle fluctuation and correlation effects. 
However, we will also point out that in other situations, the mere replacement of ``bare'' microscopic 
with effective coarse-grained rates leaves distinct renormalization effects unaccounted for and therefore
cannot fully capture the system's complex cooperative dynamics.  

The article is organized as follows: 
In section~\ref{section:Methods} the typical Monte Carlo algorithms and our notations are introduced.
Next our method for calculating the macroscopic reaction rates is outlined. 
Section~\ref{section:BirthResults} applies this technique to restricted birth processes, wherein the overall 
population growth is controlled by on-site restrictions or finite local carrying capacities. 
The ensuing macroscopic reaction rates obtained from the stochastic simulations are used to characterize 
the behavior of the system as its density restrictions become pertinent, and are compared specifically with 
the logistic model for restricted growth. 
Our method is then implemented for the well-studied models of diffusion-limited single-species pair 
coagulation and two-species binary annihilation in section~\ref{section:AnnihilationResults}. 
Applying our technique to these irreversible binary reactions confirms its consistency with the established
scaling behavior for the renormalized reaction rates.
The utility of our method is highlighted by showcasing that the macroscopic rate calculation may be 
employed to test whether the system has in fact reached the asymptotic universal scaling regime. 
Subsequently, we proceed with a study of the seminal Lotka--Volterra model for predator-prey competition
and coexistence in section~\ref{section:PredatorPreyResults}.
Here, the numerically extracted macroscopic rates are employed to probe the accuracy of the mean-field 
rate equations, utilizing the renormalized rate parameters, for modelling this system. 
Finally, we apply our technique to the rock-paper-scissors (cyclic Lotka--Volterra) and May--Leonard 
models for cyclic dominance of three competiing species in section~\ref{section:Cyclic results}, and 
investigate the modification stemming from coarse-graining the microscopic Monte Carlo parameters to 
effective macroscopic reaction rates. 
Concluding remarks and a brief summary are provided in section~\ref{section:conclusions}.

\section{\label{section:Methods} Monte Carlo simulation methods}

In this section we present the algorithm used to perform the Monte Carlo simulations, and the method of 
computing the effective, coarse-grained macroscopic reaction rates. 
The models considered in this paper comprise individual particles of multiple species $A_i$, 
$i \in \{1,2,..,S\}$, where $S$ is the total number of distinct species. 
The particles are placed on a regular cubic lattice in $d$ spatial dimensions (with $d = 2$ if not explicitly
stated otherwise), subject to periodic boundary conditions.
They may propagate diffusively via nearest-neighbor hopping, or by producing offspring on neighboring 
sites. 
Interactions are described by chemical reactions in the following manner:
\begin{equation*}
	\sum_{i=1}^S c_i^\textrm{in} A_i \xrightarrow{R_\textrm{e}} \sum_{i=1}^S 
	c_i^\textrm{out} A_i \ ,
\end{equation*}
where $c_i^\textrm{in}$ and $c_i^\textrm{out}$ denote the integer stoichiometric coefficients for species 
$A_i$ as reactants and products, respectively, in the reaction scheme.
In the lattice model, the stoichiometric coefficients $c_i^\textrm{in}$ in fact determine the specified 
conditions for a reaction to happen. 
For example, if $c_i^\textrm{in} = 2$, then the reaction can only occur if two particles meet on the same 
lattice site (``on-site" reactions), or if they encounter each other on neighboring sites (``off-site''). 
$R_\textrm{e}$ represents the reaction probability per unit time for the reaction labeled with 
$\textrm{e}$. 
Throughout this manuscript we shall use Greek letter symbols for reaction rates, 
$R_\textrm{e} = \sigma, \mu, \lambda, \ldots$. 
The microscopic reaction rates as implemented in the numerical algorithm shall be interpreted as 
propensities.

The (fictitious) time in a Monte Carlo simulation is measured in units of Monte Carlo time steps (MCS), 
which in an ecological context may be understood as approximately amounting to a generation, if all
pertinent rates are of order unity. 
For each MCS, the following algorithmic steps are performed:
\begin{itemize}
	\item Each site is assigned a propensity value which is the sum of the microscopic rates of the 		
		  particles residing in that site.
	\item A random site is picked based on its propensity value.
	\item A random individual of any species on that site is picked based on its propensity value.
	\item The chosen particle then performs one of its allowed reactions based on the propensity value of 
		  each process, {\em provided} the prescribed conditions for that process are satisfied.
	\item These steps are repeated as many times as there are number of particles present in the system
		  at that instant.
\end{itemize}

Macroscopic reaction rates are then calculated at each MCS by counting how many stochastic processes (of
a specified type) occur at this time step.
The number of reactions is then divided by the number of particles of the species performing that given 
reaction at the start of the MCS. 
Of course, one then needs to accumulate sufficient statistics over an appropriate number of independent 
simulation runs. 
We generally found that in order to obtain accurate measurements of the macroscopic rates in the 
predator-prey models of Secs.~\ref{section:PredatorPreyResults} and \ref{section:Cyclic results}, 
anywhere between twenty and one hundred independent simulation runs were sufficient. 
However, to properly determine the scaling exponents in the pair annihilation models in 
Secs.~\ref{section:CoagulationResults} and \ref{section:AnnihilationResults}, we required the statistics over $1000$ simulation runs.
The ensuing statistical error bars for our simulation data are smaller than the symbols and hence not
visible in the figures presented below.

\begin{figure*}[t]
	\includegraphics[width=0.95\columnwidth]{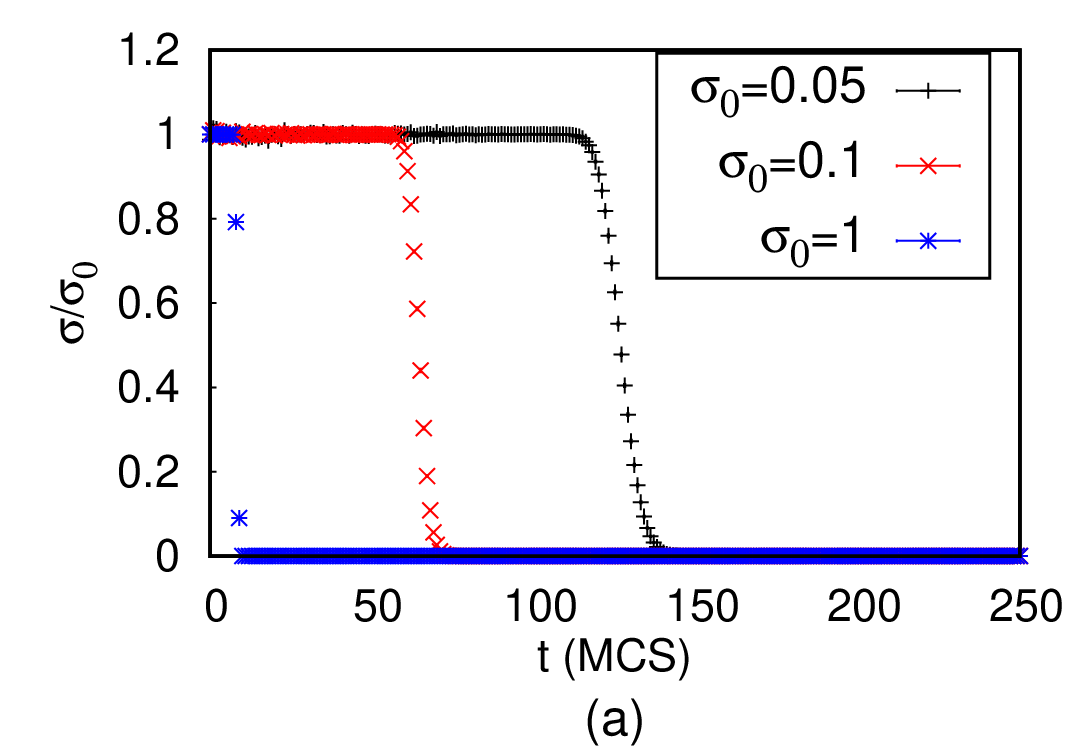} \
    	\includegraphics[width=0.95\columnwidth]{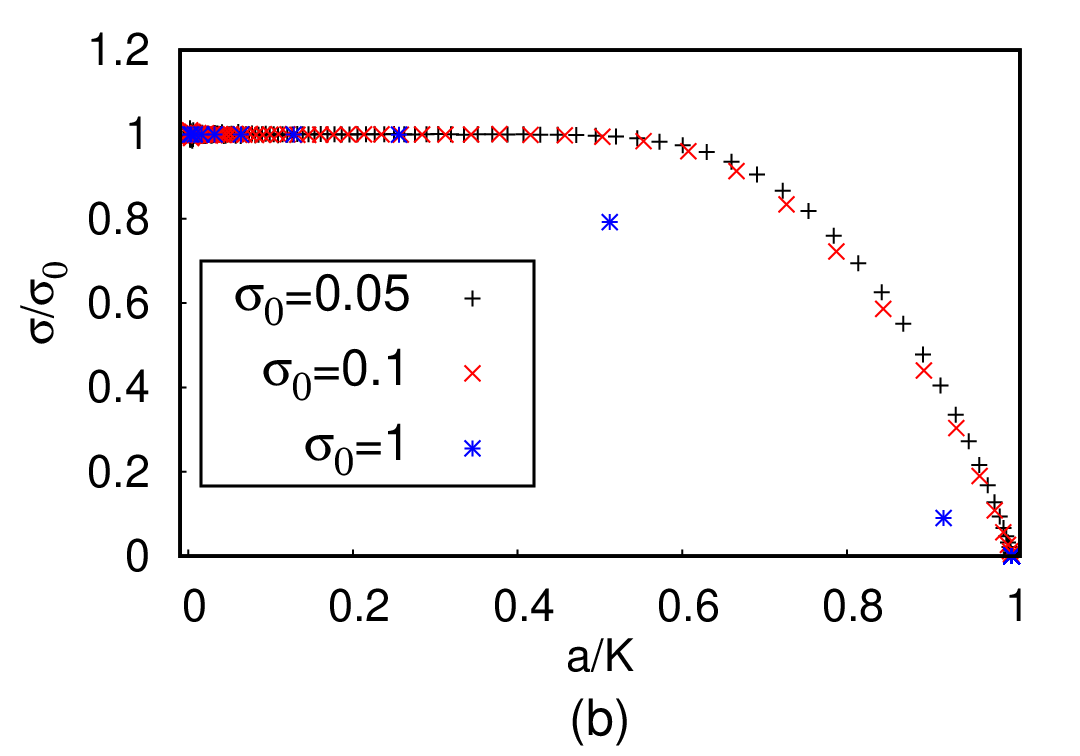}
\caption{Ratio of effective macroscopic to implemented microscopic birth rates $\sigma / \sigma_0$ for
	restricted birth processes on a square lattice with side length $L = 100$, initial
	density $a(0) = 1$, and (local) carrying capacity $K = 500$, plotted (a) as function of time $t$ (in 
	MCS) and (b) of the particle density $a / K$, for various values of $\sigma_0 = 0.05$, $0.1$, and 
	$1$, as indicated. 
	The curves were generated by averaging the data from $20$ independent simulation runs.}
\label{fig:birth_rate}
\end{figure*}
To clarify this procedure, consider the simple (unrestricted) birth reaction $A \xrightarrow{\sigma} 2 A$, 
with the particles also performing a hopping reaction $A(x_i) \xrightarrow{D} A(x_{i \pm 1})$; here 
$x_{i \pm 1}$ is to be understood to indicate one of the nearest neighbors of $x_i$ (with the lattice
constant set to unity). 
One of the allowed reactions is always attempted once a particle has been picked; i.e., a reaction only fails 
if the prescribed conditions for that reaction are not met.
Therefore, if $\sigma_0$ and $D_0$ denote the microscopic propensities of the birth and hopping 
reactions, respectively, then the probabilities for these reactions to occur are given by 
$\sigma_0 / (\sigma_0 + D_0)$ and $D_0 / (\sigma_0 + D_0)$.
Had we chosen the probabilities for each reaction such that they do not sum up to unity, then we would 
need to account for a probability $p_\phi$ for nothing to occur. 
However, one can easily show that for large systems this just results in an overall time rescaling by a 
factor $(1 - p_\phi)^{-1}$ (we have checked and confirmed this assertion in our simulations). 
The average number of birth reactions occurring per time step is then 
$\langle N_\sigma \rangle = \sigma_0 \, \langle N_A \rangle / (\sigma_0 + D_0)$. 
Furthermore, since each reaction increases the number of particles by one, and there is no other process
that alters the particle number, we can immediately write down the exact rate equation describing the time
evolution of the system:
\begin{equation*}
	\frac{d \langle N_A(t) \rangle}{dt} = \frac{\sigma_0}{\sigma_0 + D_0} \, \langle N_A(t) \rangle \ .
\end{equation*}
This confirms that mean-field rate equations become exact for {\em linear} stochastic reactions, namely
processes that only require the presence of a single particle. 
We also note that the average number of reactions occurring per time step and per particle is just 
$\sigma = \sigma_0 / (\sigma_0 + D_0)$, which could be set equal to $\sigma_0$ via rescaling time by 
the factor $(\sigma_0 + D_0)^{-1}$.

For nonlinear reactions such as $2 A \xrightarrow{\lambda} A$ (with additional hopping transport as 
before), this calculation becomes analytically intractable (beyond one dimension) due to the implied 
condition that two particles need to meet for the coagulation process to happen. 
Therefore the average number of reactions occurring per time step is $\langle N_\lambda \rangle = 
[ \lambda_0 / (\lambda_0 + D_0) ] \langle N_A \rangle \times \textrm{prob(two particles meet)}$, 
where $\lambda_0$ and $D_0$ are the microscopic propensities for pair coagulation and hopping. 
The conditional probability of two particles to meet (on site or in their immediate vicinity) cannot be 
analytically evaluated; the mean-field approximation resides in the assumption of a uniform density of 
$A$ particles, such that $\textrm{prob(two particles meet)} \propto a$, where $a$ is the mean $A$ 
density; this directly leads to the standard coagulation rate equation
\begin{equation}
	\frac{d\,a(t)}{dt} \approx - \lambda_0 \, a(t)^2 \ .
\label{eq:coagmf}
\end{equation}

In this study, all models are simulated on regular cubic lattices with equal side lengths $L$.
We apply periodic boundary conditions, i.e., set a toroidal topology to eliminate boundary effects. 
As mentioned above, the input parameters for the Monte Carlo simulation are the prescribed microscopic 
rates, measured in units of inverse MCS, and hence the probability for a reaction to occur is given by the 
ratio of that rate to the sum of the rates for all possible processes.
If nearest-neighbor hopping is implemented, it is always chosen to take place, with a microscopic rate 
$D_0 = 1$.

\section{\label{section:BirthResults} Site-restricted birth processes}

In this section we first consider the simple example of a single species of particles that can diffuse and 
reproduce on a two-dimensional square lattice. 
The nonlinearity of the system is introduced via an on-site restriction parameterized through the local 
carrying capacity $K$, which represents the maximum number of particles allowed at each site. 
Therefore the indidivuals undergo branching reactions $A \xrightarrow{\sigma} 2 A$, where the offspring 
are generated on a randomly picked nearest-neighbor selected from the von-Neumann neighborhood.
Birth reactions succeed with a certain microscopic probability $p_\sigma$, but only if the chosen adjacent
site holds a particle occupation number below the carrying capacity $n \leq K$. 
We are interested in measuring the macroscopic reaction rate $\sigma$, which we define as the number of
reactions occurring per MCS and per particle: 
Thus, we take $\sigma$ to be the coarse-grained per-capita reaction rate that effectively incorporates
emerging correlations induced by successively filling the lattice sites as time progresses and the overall
particle density increases. 
Consequently, the macroscopic reaction rate will become reduced as function of time and density.
This rather elementary example illustrates the usefulness of Monte Carlo simulations to determine 
effective macroscopic rates, capturing nontrivial correlation effects.

\begin{figure*}[t]
	\includegraphics[width=0.95\columnwidth]{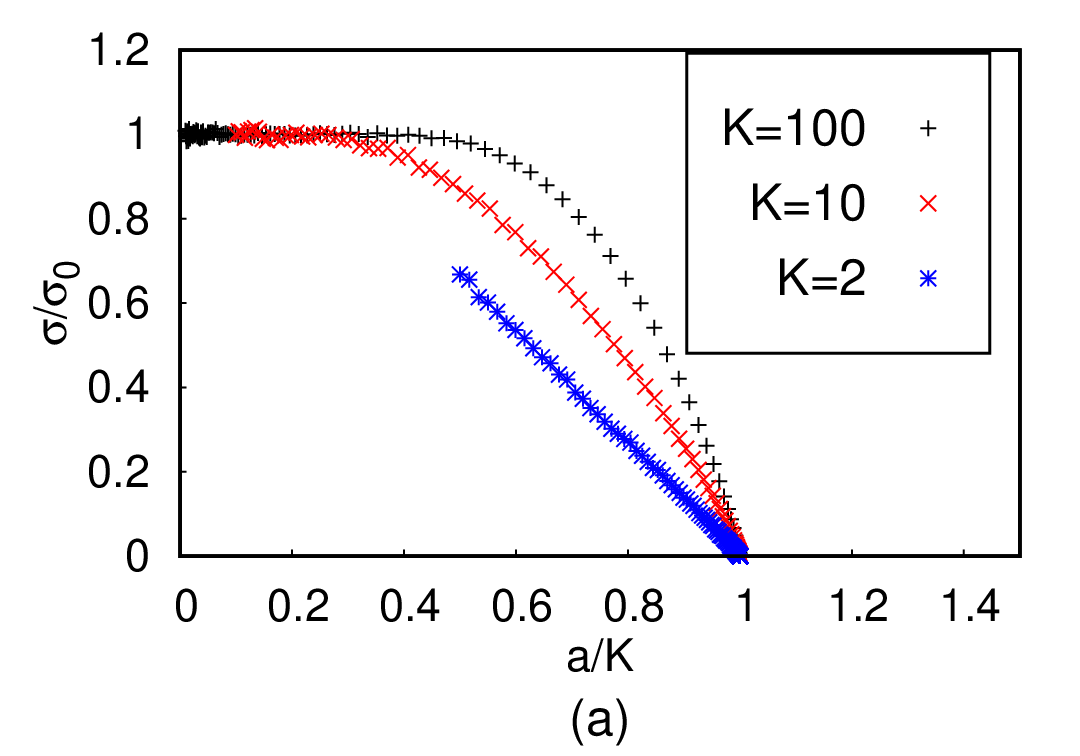} \
	\includegraphics[width=0.95\columnwidth]{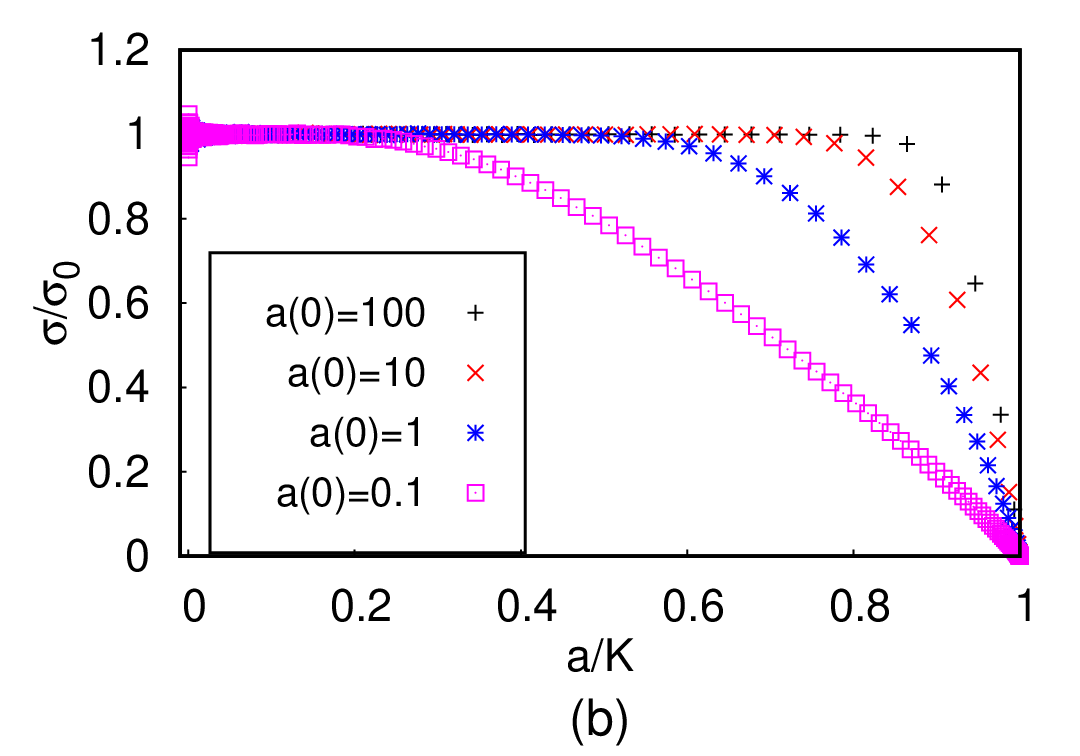} 
\caption{Ratio of macroscopic and microscopic birth rates $\sigma / \sigma_0$ for restricted birth 
	processes on a square lattice with side length $L = 100$ and $\sigma_0 = 0.05$ as function of the
	density $a / K$, for (a) fixed initial density $a(0) = 1$ and various values of carrying capacity (site 	
	occupation restriction) $K = 100$, $10$, and $2$; (b) fixed carrying capacity $K = 500$ and various 		initial densities $a(0) = 100$, $10$, $1$, and $0.1$. 
	The data were averaged over $20$ independent simulation runs.}
\label{fig:birth_rate_K}
\end{figure*}
Fig.~\ref{fig:birth_rate}(a) shows the effective macroscopic rate $\sigma$ relative to the microscopic 
birth rate $\sigma_0$ as a function of time $t$. 
Of course, initially, when the particle density is low, the macroscopic rate is equal to the microscopic one,
$\sigma(t = 0) = \sigma_0$. 
Yet once the lattice starts filling up, we observe a rapid decrease in the macroscopic rate $\sigma$. 
The time it takes to reach this point diminishes for larger microscopic $\sigma_0$. 
These results indicate that the system can be described through unconstrained linear birth kinetics right
until the lattice becomes almost saturated. 
Since the crossover between these two regimes is quite abrupt, we can extract a well-defined characteristic 
time scale $t_c$ for the lattice to fill up and the local carrying capacity restrictions to impede further
growth.
Later in this section, we shall describe a more convenient method to compute this relevant time scale for
the macroscopic dynamics. 
The coarse-grained effective rate as a function of the increasing particle density $a$ is displayed in 
Fig.~\ref{fig:birth_rate}(b). 
This plot demonstrates that while $\sigma(a = 0) = \sigma_0$ for low densities, the macroscopic birth 
rate $\sigma$ decreases drastically with the mean density $a$, as one would expect.
We note that the graphs for low microscopic rates $\sigma_0 = 0.05$ and $0.1$ are very close, and the 
correlation-induced reduction of the macroscopic rate begins at almost the same filling fraction 
$a / K \geq 0.6$. 
We found the decrease in $\sigma(a / K)$ to not be described by a simple exponential: 
A log-linear plot of this function does not yield a straight line.

To gain a better understanding of the kinetics of this system, in Fig.~\ref{fig:birth_rate_K} we plot the 
macroscopic rate $\sigma / \sigma_0$ against the lattice filling ratio $a / K$, as in 
Fig.~\ref{fig:birth_rate}(b), but now for different values of (a) the carrying capacity $K$ and (b) the initial
densities $a(0)$. 
These results show that the filling fraction at which the macroscopic rate $\sigma$ begins to markedly
diminish depends nontrivially on $K$. 
Futhermore, the function $\sigma(a)$ does not simply follow the standard linear logistic model behavior, 
according to which $\sigma(a) = \sigma_0 \left( 1 - a / K \right)$.
Only asymptotically, as the density approaches the carrying capacity, one may apply a linear fit to the
macroscopic rate as function of $a$.
For low carrying capacity $K = 2$, our chosen initial density $a(0) = 1$ implies that the system is 
half-filled at the outset of the simulation.
Indeed, in this case the asymptotic linear regime has already been reached, and a straightforward data fit
yields $\sigma(a) = 0.067 - 0.033\, a$.
Consequently, the effective birth rate becomes renormalized from $\sigma_0 = 0.05$ to 
$\sigma_r = 0.067$. 
In contrast, the carrying capacity does not acquire any renormalization, and the slope of this linear fit in 
fact equals half of the y-intercept, as predicted by the logistic equation with $K = 2$. 
We found this to be a general feature for the long-time asymptotic behavior for other $K$ values as well,
as will be discussed below.

The growth function in the logistic model does not depend on the initial density. 
This is not true for our lattice birth model with on-site restriction, as demonstrated in 
Fig.~\ref{fig:birth_rate_K}(b). 
Simulations with larger values of $a(0)$ show that the macroscopic rate $\sigma$ only begins to decrease 
at higher lattice fillings. 
While this observation might appear counter-intuitive at first glance, it is due to the initialization of the
simulation runs with a Poissonian particle number distribution. 
Since the different sites are initially uncorrelated, it takes some time for mutual correlations to develop, 
and for some lattice sites to reach maximum capacity. 
Hence, by the time that a simulation with an initially lower $a(0)$ value reaches a high overall density $a$,
some sites are already full, and the growth is restricted to the boundaries of filled zones.
Indeed, if we had instead graphed the time-dependent macroscopic birth rate $\sigma(t)$, the systems 
initialized with higher $a(0)$ start filling up sooner, as one would expect.

\begin{figure}[t]
	\includegraphics[width=0.95\columnwidth]{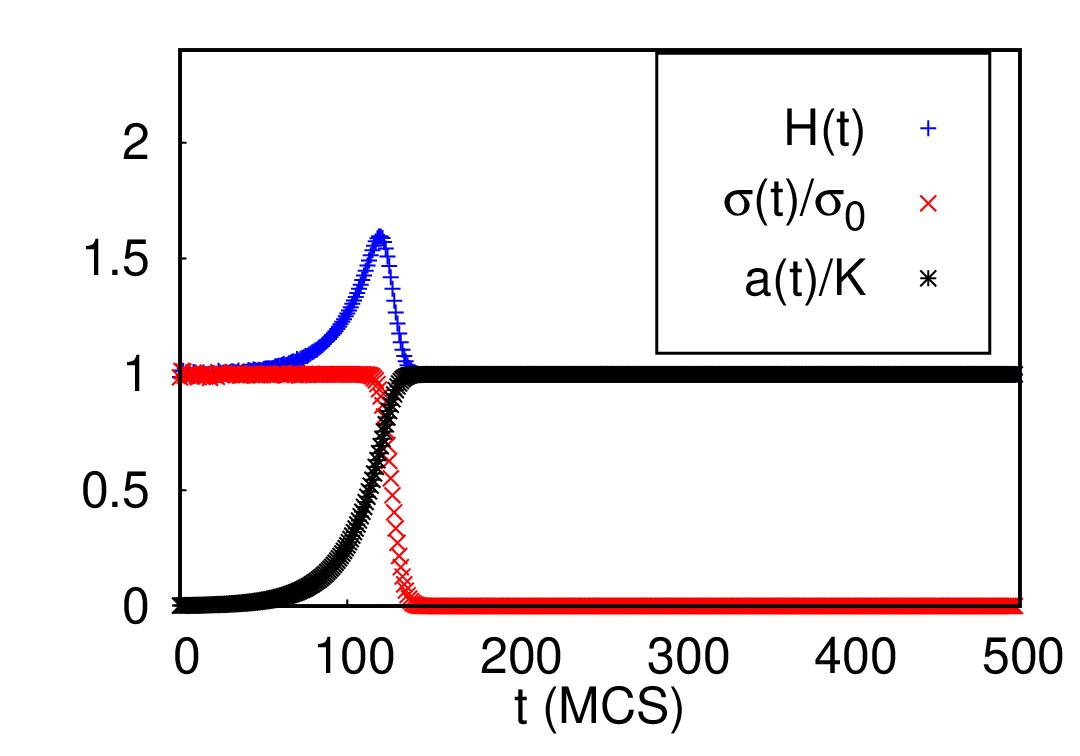} 
\caption{The functions $H(t) = a(t) / K + \sigma(t) / \sigma_0$, $\sigma(t) / \sigma_0$, and $a(t) / K$ 
	vs. time $t$ (in MCS) for restricted birth processes on a square lattice with side length $L = 100$, 
	initial density $a(0) = 1$, birth rate $\sigma_0 = 0.05$, and carrying capacity $K = 500$. 
	The data were averaged over $20$ independent simulation runs.}
\label{fig:birth_conserved} 
\end{figure}
\begin{figure}[b]
	\includegraphics[width=0.95\columnwidth]{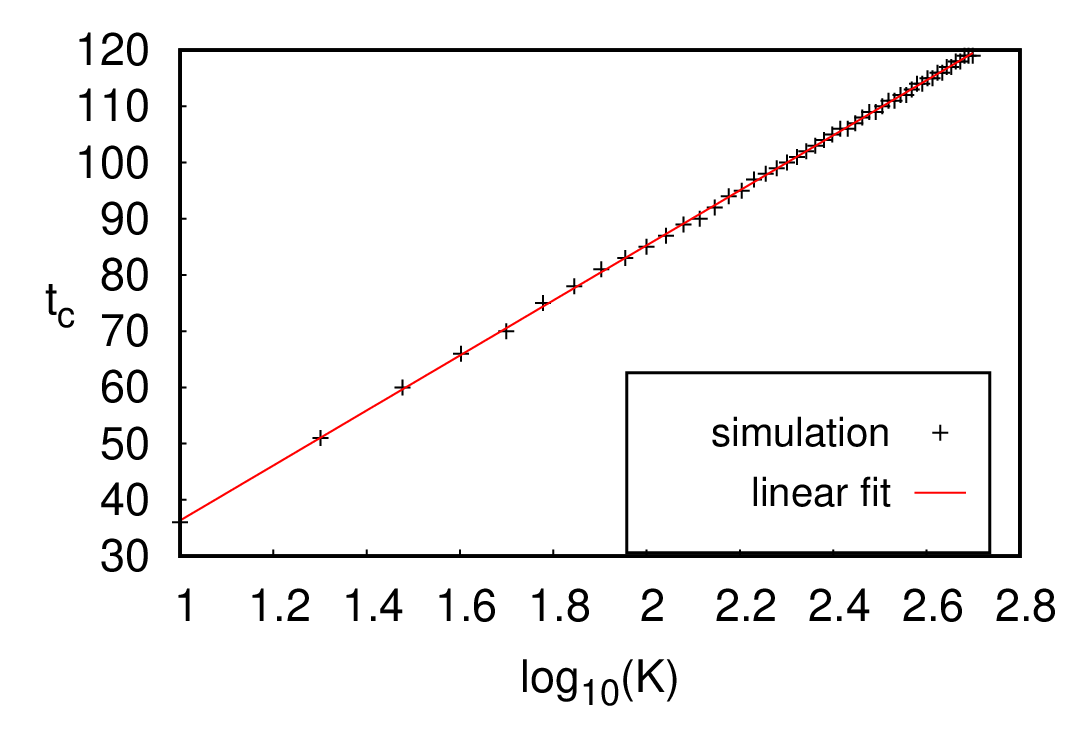} 
\caption{Characteristic crossover time $t_c$ for the lattice to become saturated vs. carrying capacity 
	$\log_{10} K$ in the restricted birth model on a two-dimensional square lattice with side length $L = 100$, 
	initial density $a(0) = 1$, and $\sigma_0 = 0.05$. 
	The data were averaged over $20$ independent simulation runs.}
\label{fig:birth_t_c_K}
\end{figure}

Measuring the macroscopic birth rate $\sigma$ can be utilized to accurately compute the characteristic 
time $t_c$ it takes the system to be affected by the on-site restrictions. 
This method relies on the fact that the logistic model features a ``conserved'' quantity, 
$H(t) = a(t) / K + \sigma(t) / \sigma_0 = 1$. 
However, $H(t)$ is no longer constant in time in the lattice model; as shown in 
Fig.~\ref{fig:birth_conserved}, this quantity reaches a maximum when the macroscopic rate starts to
decrease as a consequence of emerging local occupation restrictions and correlations. 
Hence, locating the maximum of $H(t)$ in time provides a refined scheme to calculate $t_c$. 
Applying this prescription, the characteristic crossover time $t_c$ for the lattice to fill up is plotted against 
the carrying capacity $K$ in Fig.~\ref{fig:birth_t_c_K}. 
Our measurements yield a logarithmic dependence of $t_c$ with respect to $K$, which indicates that the 
mean density $a$ grows exponentially up to $t_c$. 

Furthermore, the density at $t = t_c$ sets a characteristic value $a_c = a(t_c)$ at which the birth
restrictions in the appreciably filled lattice begin to affect the growth kinetics.
One might hypothesize that $a_c$ represents a universal filling fraction, regardless of the system 
parameters. 
However, investigating the dependence of $a_c$ on $K$ demonstrates that $a_c$ also depends 
logarithmically on $K$. 
This can be explained by noting that for higher $K$ values, the local occupation numbers may grow on 
more sites before the on-site restrictions take effect. 
Therefore, by the time that a single site has reached maximum occupancy, the density correlations have 
already expanded to a larger distance than would have been possible for smaller carrying capacity. 
Hence, at the instant $t_c$ when $H(t)$ reaches its maximum, the typical lattice filling will be higher for 
larger values of $K$. 
In summary, the initial kinetics of the restricted birth model can be characterized as follows: 
The density $a$ grows exponentially with time, until it reaches the critical lattice filling $a_c / K$, beyond 
which the growth rate rapidly drops to zero. 
This critical lattice filling ratio is logarithmically dependent on the prescribed carrying capacity $K$.

\begin{figure}[t]
    	\includegraphics[width=0.95\columnwidth]{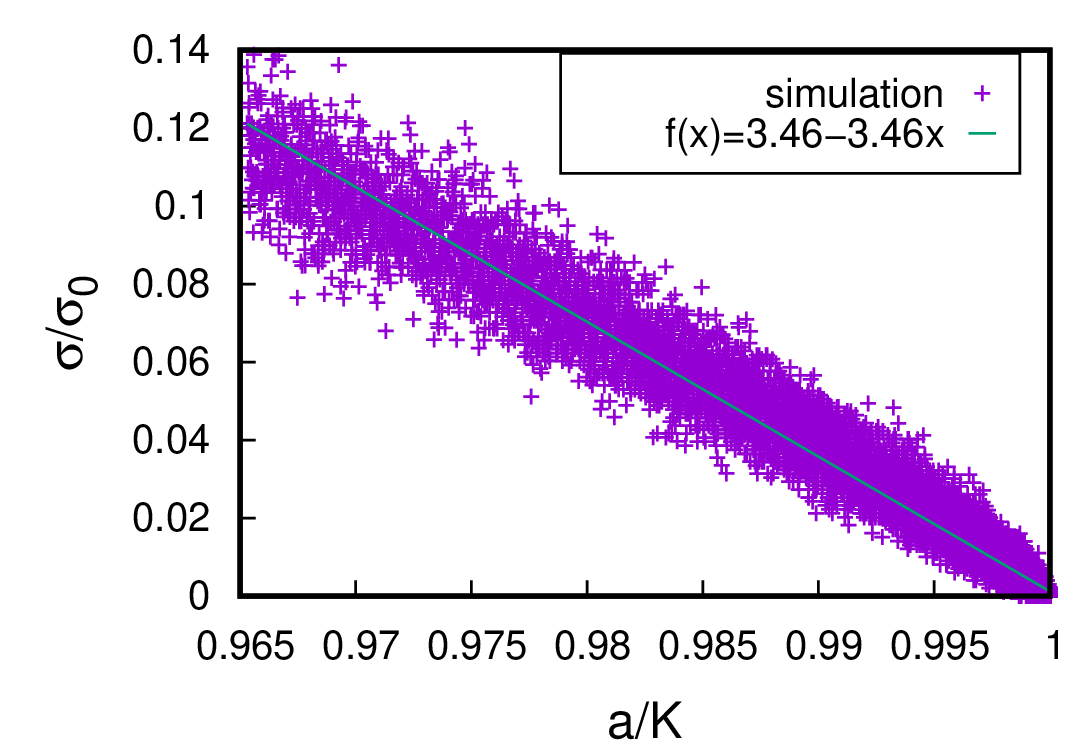} 
\caption{Asymptotic long-time behavior of the ratio of macroscopic and microscopic birth rates 
	$\sigma / \sigma_0$ as function of the density $a / K$ for restricted birth processes on a square 	
	lattice with side length $L = 100$, initial density $a(0) = 1$, small birth rate $\sigma_0 = 0.0001$, 
	and carrying capacity $K = 10$. 
	The data were averaged over $100$ independent simulation runs.}
\label{fig:birth_asymptotic}
\end{figure}
In order to probe the long-time behavior of this system, we ran simulations with small birth rates 
$\sigma_0$, such that enough data points at the end of the simulation were available. 
Figure~\ref{fig:birth_asymptotic} confirms that the asymptotic regime can be modeled using a decreasing 
linear fit for the function $\sigma(a)$, which matches the logistic rate equation. 
The fact that the appropriate linear fit function has equal $y$-intercept and slope suggests that the 
emerging effective carrying capacity value is identical with the implemeted on-site restriction $K$. 
In contrast, the reaction rate $\sigma_r = Z \, \sigma_0$ attains a multiplicative renormalization factor 
$Z$. 
Therefore, the asymptotic long-time kinetics for the restricted birth processes can be represented well by
$da / dt = Z \, \sigma_0 \, a \left( 1 - a / K \right)$, where $K$ is the microscopic on-site restriction.
We have indeed tested and confirmed this hypothesis for different $K$ values.

\begin{figure*}[t]
    	\includegraphics[width=0.66\columnwidth]{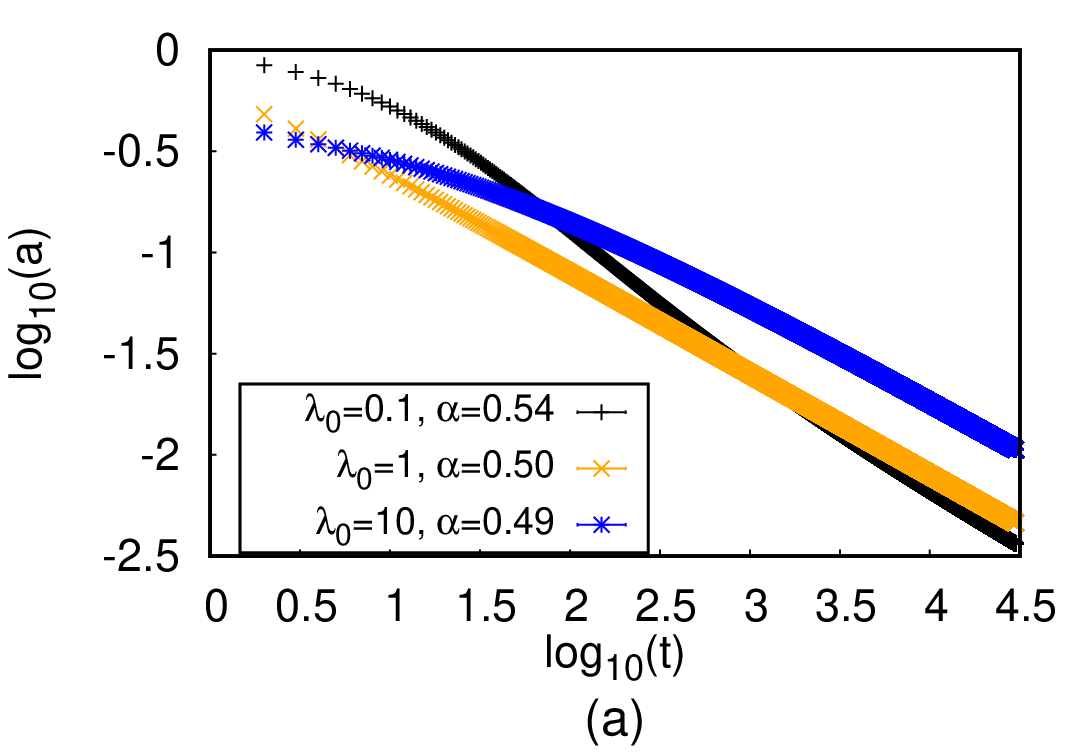} \
    	\includegraphics[width=0.66\columnwidth]{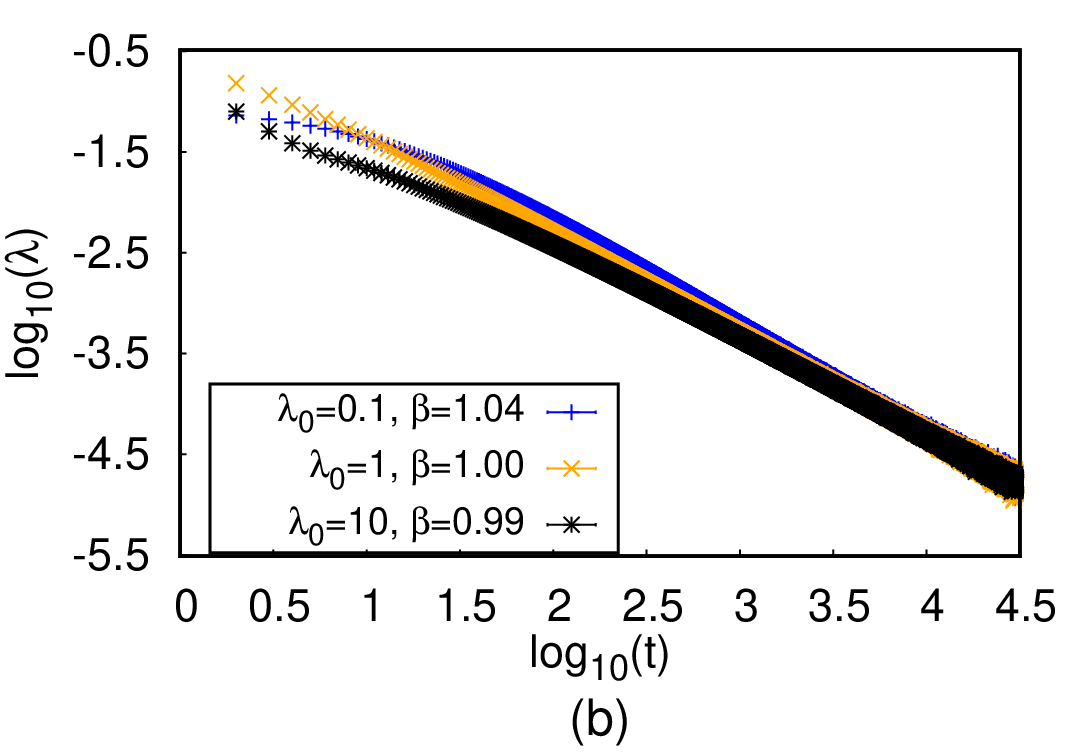} \
    	\includegraphics[width=0.66\columnwidth]{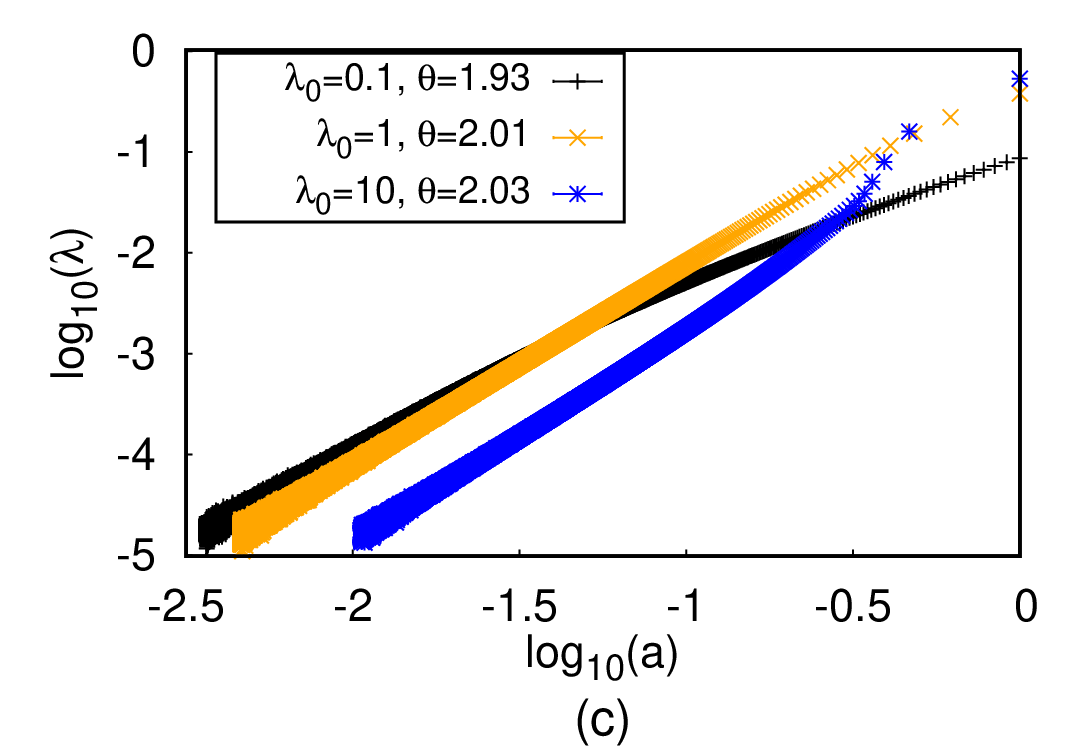} 
\caption{Single-species pair coagulation kinetics $2 A \xrightarrow{\lambda} A$ on a one-dimensional 
	lattice of size $L = 1,500,000$ with initial density $a(0) = 1$, for different microscopic reaction rates 
	$\lambda_0 = 0.1$, $1$, and $10$, as indicated: 
	Double-logarithmic plots of (a) the particle density decay with time $a(t)$; the macroscopic effective 
	reaction rate (b) as function of time $\lambda(t)$, and (c) as function of density $\lambda(a)$.
	The resulting effective scaling exponents $\alpha$, $\beta$, and $\theta$ are listed.
	The data were averaged over $1000$ independent runs.}
\label{fig:cog_1d}
\end{figure*}
For $t < t_c$, we found the growing density to be nicely fitted by an exponential $a(t) \sim e^{r \, t}$, 
with a value for the parameter $r$ that is close to $\sigma_0$. 
This allows us to fully characterize the restricted birth model as follows: 
The density $a$ grows exponentially with the microscopic rate value $\sigma_0$ for $t < t_c$. 
In the asymptotic region $t \gg t_c$, the dynamics can be described by a logistic rate equation with 
renormalized effective rate $Z \sigma_0$. 
Near $t \gtrsim t_c$ the kinetics interpolates between these two distinct regimes.

\section{\label{section:AnnihilationResults} Diffusion-limited binary annihilation reactions}

We next address measuring the drastic rate renormalizations induced by the emerging intrinsic dynamical 
correlations in the diffusion-limited binary reaction processes $A + A \to A$ and $A + B \to \emptyset$.

\begin{figure*}[t]
    	\includegraphics[width=0.66\columnwidth]{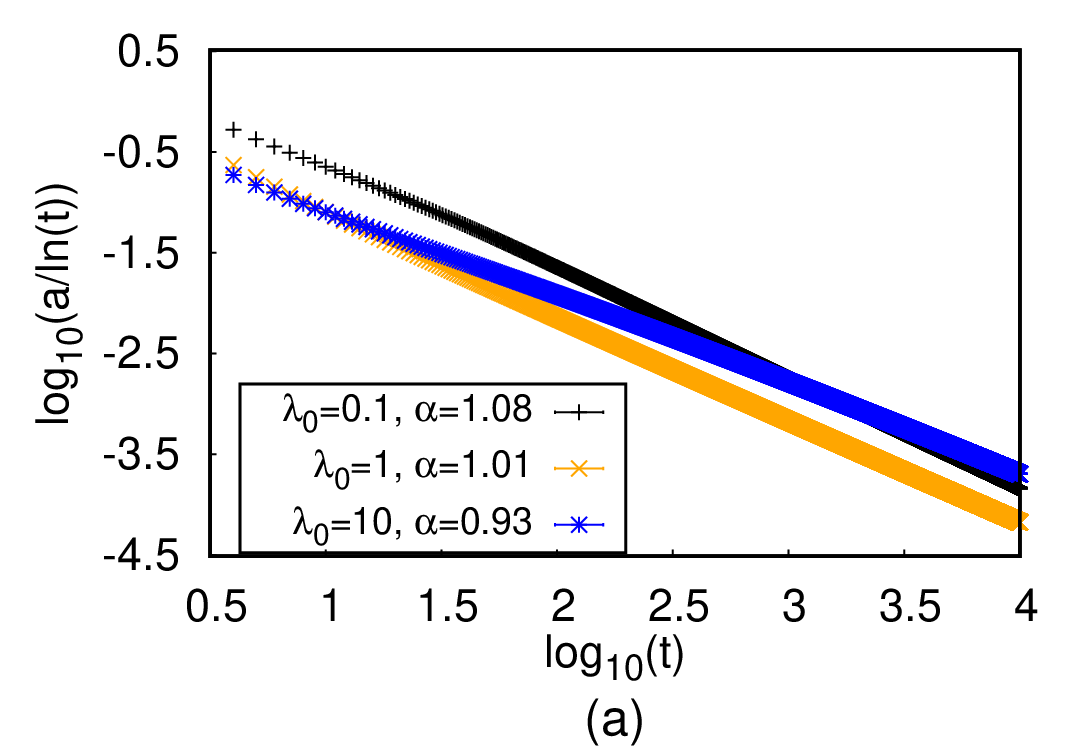} \
   	\includegraphics[width=0.66\columnwidth]{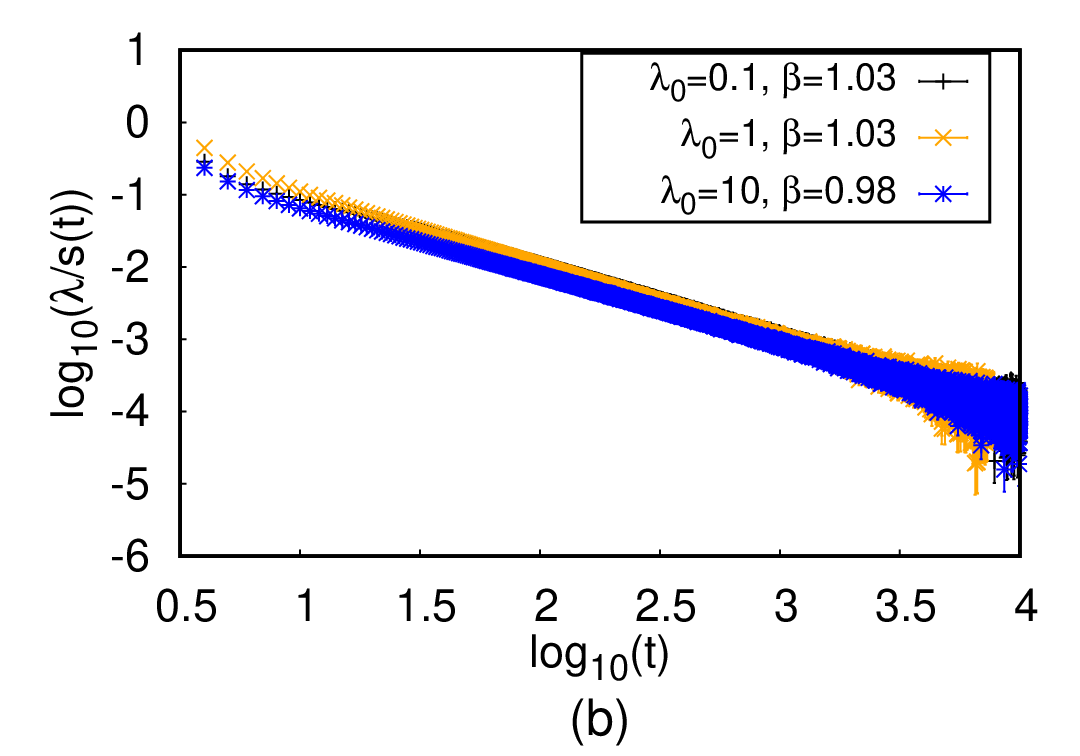} 
\caption{Single-species pair coagulation kinetics $2 A \xrightarrow{\lambda} A$ on a two-dimensional 	
	square lattice of linear size $L = 256$ with initial density $a(0) = 1$, for different microscopic 
	reaction rates $\lambda_0 = 0.1$, $1$, and $10$, as indicated: 
	Double-logarithmic plots of the time dependence of (a) the particle density $a(t) / \ln{t}$ and (b) the 
	macroscopic effective reaction rate $\lambda(t) / s(t)$, with $ s(t) = \left( 1 - \alpha / \ln{t} \right)$, 
	where the value of $\alpha$ obtained from the simulations was used.
	The resulting effective scaling exponents $\alpha$ and $\beta$ are listed.
	The data were averaged over $1000$ independent simulation runs.}
\label{fig:cog_2d}
\end{figure*}
\subsection{\label{section:CoagulationResults} Single-species pair coagulation}

In dimensions below the upper critical dimension, $d \leq d_c = 2$, the density decay in the single-species 
binary coagulation reaction $2 A \xrightarrow{\lambda} A$ with diffusively spreading particles is known to 
deviate from the mean-field scaling $a(t) \sim t^{-\alpha}$ with $\alpha = 1$, which directly follows from 
the rate equation (\ref{eq:coagmf}) \cite{EffectiveRateEquation, CogDoi, CogPeliti, Lee2, ft2, Cog1, Cog2, 
Cog3, Cog4, Cog5, CogAnn6, ft10}.
In the long-time asymptotic region, the system is then no longer controlled by the reactivity 
$\propto \lambda$, but by the gradually diminishing chance for pairs of particles to meet.
In this asymptotic diffusion-limited regime, the characteristic distance scale is set by the diffusion length
$\ell_D(t) \sim (D \, t)^{1/2}$ with diffusivity $D$, whence the long-time decay of the density 
$a \sim \ell_D^{-d}$ is governed instead by the exponent $\alpha = d / 2$ \cite{CogAnn1, CogAnn2, 
CogAnn3, CogAnn4, CogAnn5, CogAnn6, CogAnn7}.
In this paper, we also specifically focus on the scaling of the effective macroscopic coagulation rate with 
time $\lambda(t) \sim t^{- \beta}$ and the mean particle density $\lambda(a) \sim a^{\theta}$. 
The scaling exponents $\alpha$ and $\theta$ are intimately related, while $\beta$ is fixed by the definition 
of the macroscopic reaction rate $\lambda$. 
Therefore a single independent scaling exponent fully characterizes the diffusion-limited pair coagulation 
model, which we shall employ to showcase the applicability of our algorithm in determining the strongly 
renormalized scaling of the macroscopic reaction rate and its consistency with the correct asymptotic 
density decay.

Before discussing the detailed simulation results, we address the relationship between the exponents 
$\alpha$ and $\theta$, and determine the value of the exponent $\beta$. 
As mentioned in Sec.~\ref{section:Methods}, we define the macroscopic per-capita reaction rate as the 
number of reactions occurring per time step, relative to number of particles then present in the system.
Thus, we can write the temporal evolution of the mean particle density as an ordinary differential equation 
in the following manner~\footnote{It is common in the literature to define 
$\frac{da}{dt} = - \tilde{\lambda}(a) \, a^2$. 
However, in this case $\tilde{\lambda}(a)$ should be more adequately viewed as a running coupling 
rather than a macroscopic reaction rate.},
\begin{equation}
	\frac{da}{dt} = - \lambda(a) \, a \ ,
\label{eq:coagre}
\end{equation}
where the nontrivial function $\lambda(a)$ incorporates fluctuation and correlation effects on a 
macroscopic scale.

Consequently, the coarse-grained per-capita reaction rate is $\lambda = - d \ln{a} / dt$; with 
$a(t) \sim t^{- \alpha}$, we obtain $\lambda \sim \alpha / t \sim a^{1 / \alpha}$, and hence
\begin{equation}
	\beta = 1 \ , \quad \theta =  1 / \alpha \ .
\label{eq:scarel}
\end{equation}
The exponents $\alpha$ and $\theta$ describe the asymptotic correlated dynamics of the system, whereas 
measuring the exponent $\beta$ is indicative of deviations from the scaling regime, namely when 
$\beta \not= 1$. 
The mean-field rate equation approximation predicts $\alpha = \theta = 1$, whereas an exact analysis of 
the long-time asymptotic scaling gives $\alpha = d / 2$ and therefore $\theta = 2 / d$ for $d < d_c = 2$. 
Above the critical dimension, the mean-field power laws are recovered; while at $d_c$, the mean-field 
scaling is modified by logarithmic corrections. 
In the following, we shall address the kinetics in dimensions $d = 1$, $2$, and $3$ separately.

For our numerical investigations of this model, we considered two distinct algorithmic variants, namely
(i) an on-site implementation where multiple site occupancy is permitted, and the coagulation reaction 
occurs if two particles meet on the same lattice site; and (ii) an off-site implementation where a maximum
occupancy of one particle per site is enforced, and the reaction occurs if two particles encounter each other 
on nearest-neighbor sites. 
For densities much bigger than unity, the on-site implementation behaves differently than the off-site
variant, which follows the above description:
When many particles are present on the same location, the condition for a coagulation reaction to happen 
is actually always satisfied, which means that the system is effectively subject to the single-particle death
reactions $A \to \emptyset$ rather than binary processes, leading to exponentially fast temporal decay.
Hence, in this work we restrict ourselves to the off-site implementation which does not display such 
algorithmic artifacts even at high densities.

\begin{figure*}[t]
	\includegraphics[width=0.66\columnwidth]{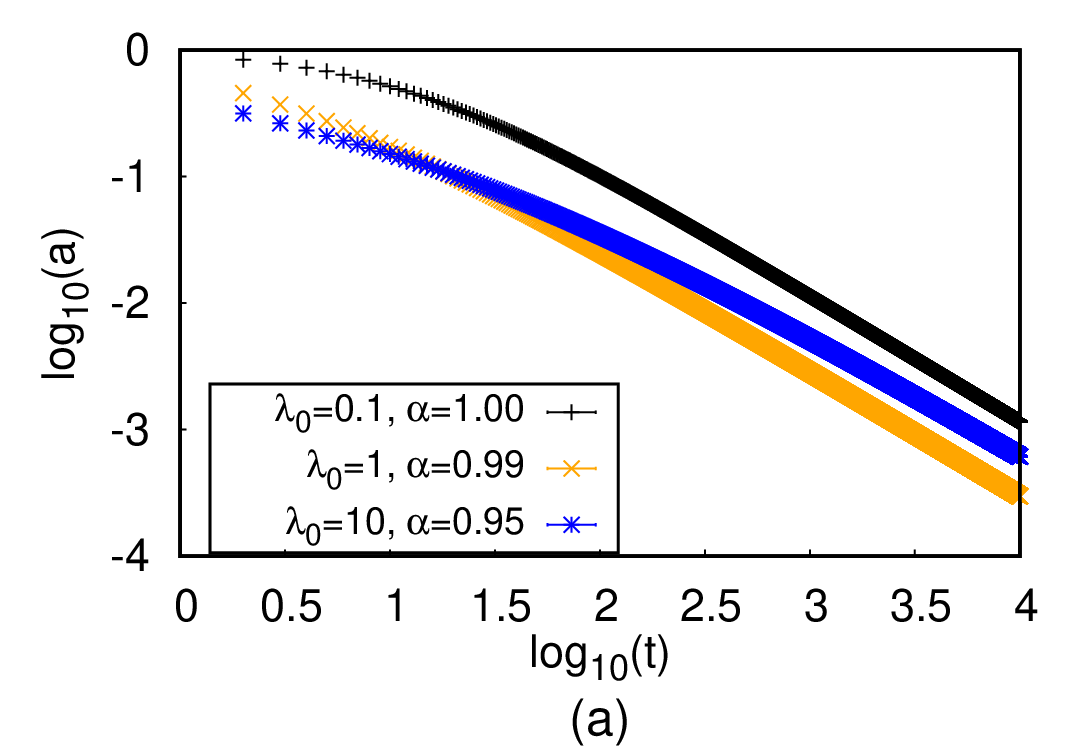} \
    	\includegraphics[width=0.66\columnwidth]{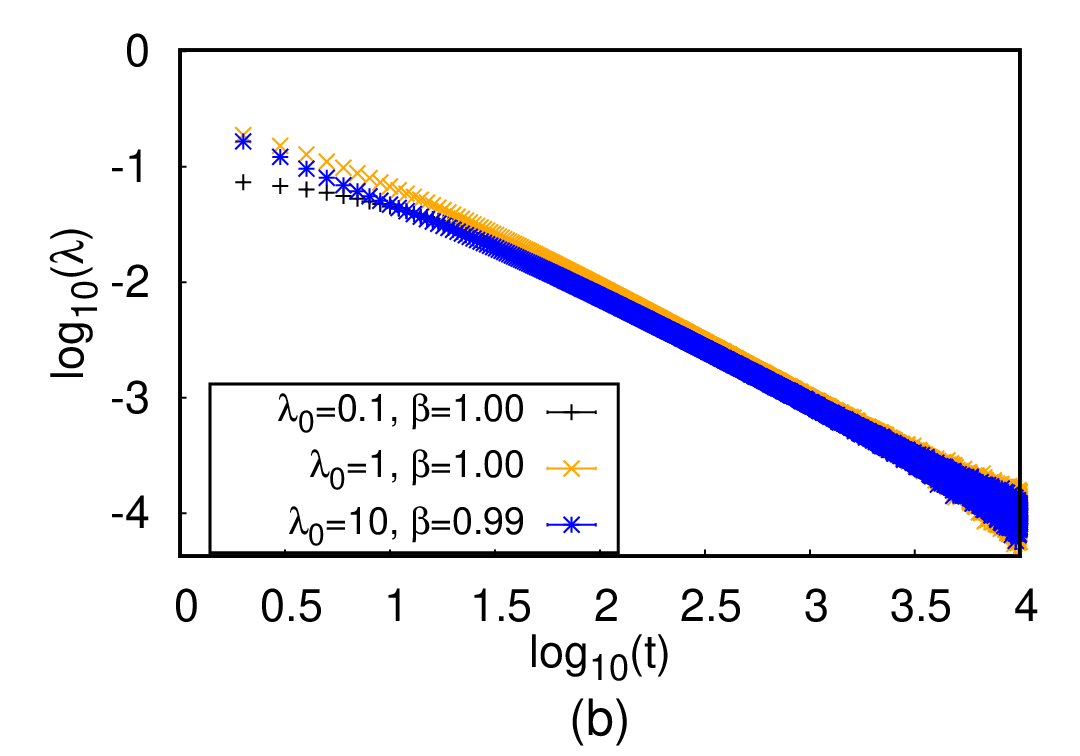} \
	\includegraphics[width=0.66\columnwidth]{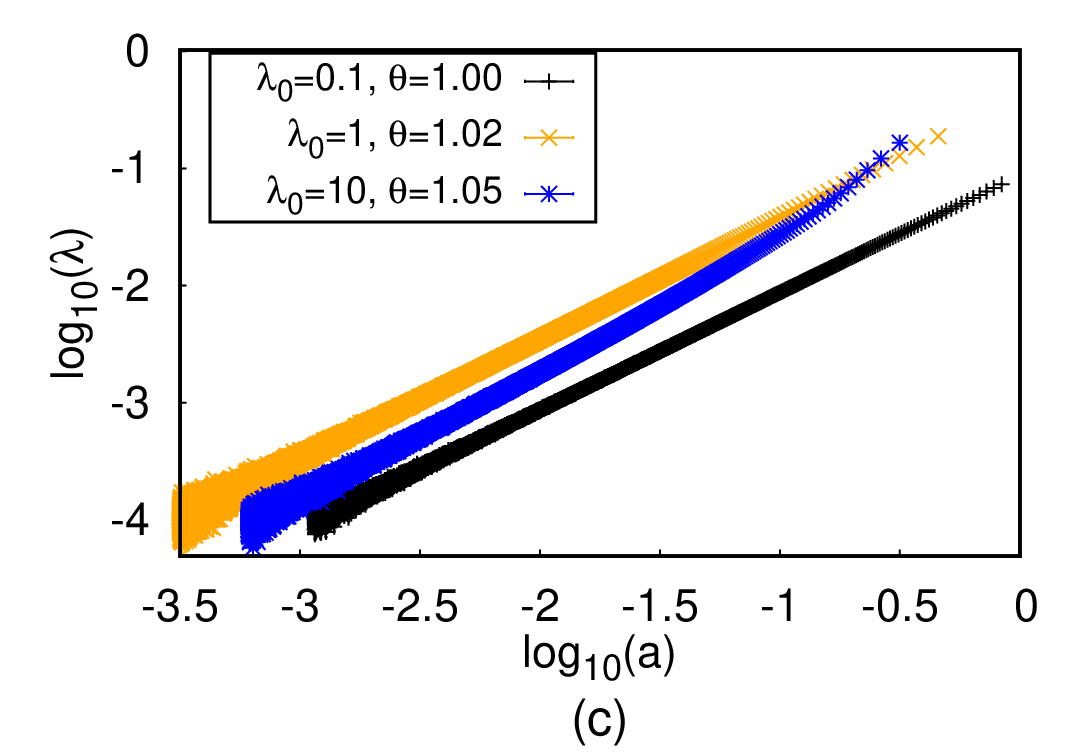} 
\caption{Single-species pair coagulation kinetics on a three-dimensional cubic lattice of linear size 
	$L = 100$ with initial density $a(0) = 1$, for different microscopic reaction rates 
	$\lambda_0 = 0.1$, $1$, and $10$, as indicated: 
	Double-logarithmic plots of (a) the particle density decay with time $a(t)$; the macroscopic effective 
	reaction rate (b) as function of time $\lambda(t)$, and (c) as function of density $\lambda(a)$.
	The resulting effective scaling exponents $\alpha$, $\beta$, and $\theta$ are listed.
	The data were averaged over $1000$ independent runs.}
\label{fig:cog_3d}
\end{figure*}
For a large one-dimensional closed chain, Fig.~\ref{fig:cog_1d} depicts double-logarithmic graphs for the 
temporal decay of the  mean particle density $a(t)$ and the time- and density dependences of the effective 
macroscopic coagulation rate $\lambda$, which allows us to measure the three associated scaling 
exponents $\alpha$, $\beta$, and $\theta$.
These power laws were numerically determined by applying linear fits to the double-logarithmic graphs in
the long-time asymptotic region; the latter was determined via appropriately discarding early-time data 
that produced an effective exponent $\beta \not= 1$ and retaining the late-time regime where $\beta$ 
was closest to unity.
We have performed this analysis here for $\lambda_0$ values in the reaction-limited regime 
($\lambda_0 = 0.1$), deep in the diffusion-limited regime ($\lambda_0 = 10$), and at the intermediate 
scale, where nearest-neighbor hops and binary reactions are implemented with the same probability 
($\lambda_0 = 1$). 
Simple mean-field scaling can obviously not adequately describe the coagualation kinetics.
For $\lambda_0 = 1$, we measure $\beta = 1$ already after $100$ MCS, indicating that the system has
perfectly reached the scaling regime.
Correspondingly, we find clean values $\alpha = 1/2$ and $\theta = 2$ adequate for diffusion-limited 
scaling with a strongly renormalized reaction rate, indicative of strong particle anti-correlations and the
presence of depletion zones.
For small $\lambda_0$, we mostly observe the reaction-limited regime where the particles remain rather 
well-mixed, generating effective exponents $\alpha$ and $\theta$ in our data that approach the 
diffusion-limited regime, but are still situated between the mean-field predictions $\alpha = \theta = 1$ 
and the above asymptotic values.  
If an adequately large lattice were afforded sufficient longer simulation time, the kinetics would ultimately 
cross over fully to the diffusion-limited regime. 
However, for $\lambda_0 = 10$, the effective scaling exponents determined during our simulation time
window also differ slightly from the expected numbers, but are of course well distinct from the rate 
equation predictions. 
For such extreme discrepancies between hopping and reaction rates, one would have to run the simulation
significantly longer to arrive at more reliable estimates for the asymptotic scaling exponents. 

Two dimensions constitutes a special situation for diffusion-controlled pair coagulation as the boundary 
dimension between mean-field scaling with $\alpha = \theta = 1$ for $d > d_c = 2$ and strongly 
renormalized scaling exponents $\alpha = 1 / \theta = d / 2$ for $d < d_c$.
At $d_c = 2$, the mean-field scaling behavior becomes modified by logarithmic corrections: 
According to the renormalization group analysis, the particle density in the long-time asymptotic regime 
decays according to $a(t) \sim \ln{t} / t^\alpha$ (with $\alpha = 1$ here) that in turn implies 
$\lambda(t) \sim \left( 1 - \alpha / \ln{t} \right) / t$. 
One could also compute $\lambda(a)$ in terms of the Lambert $W$ function, yet this results in a 
complicated functional dependence, which would be difficult to accurately confirm in Monte Carlo 
simulation data.
The density scaling for pair coagulation at $d_c = 2$ is explored in Fig.~\ref{fig:cog_2d} for a square
lattice. 
The density decay exponent $a(t) / \ln{t} \sim t^{-\alpha}$ is found to be close to the predicted value
$\alpha = 1$; as in $d = 1$, it shifts towards smaller values for $\lambda_0 = 10$. 
As for the effective renormalized reaction rate decay $\lambda(t)$, we discern small deviations from the 
expected behavior, which indicates that prevalent internal reaction noise in our finite system precludes perfect agreement of the simulation data with the anticipated asymptotic scaling.

The mean-field density decay $a(t) \sim 1 / t$ that follows from the rate equation (\ref{eq:coagmf}) is 
recovered for a three-dimensional cubic lattice, since for $d > d_c = 2$ fluctuations do not alter the 
scaling exponents. 
Rather, stochastic fluctuations and still present particle anti-correlations merely renormalize the 
microscopic parameters implemented in the Monte Carlo simulation. 
Figure~\ref{fig:cog_3d} demonstrates that the scaling behavior is indeed governed by the mean-field 
eponents $\alpha = \beta = \theta = 1$, since their measured values nicely agree with these predictions. 
As suggested by the data shown in Fig.~\ref{fig:cog_3d}(b), the closer the system is to the asymptotic
scaling regime, the nearer the associated exponent values are to unity.

\begin{figure*}[t]
    \includegraphics[width=0.66\columnwidth]{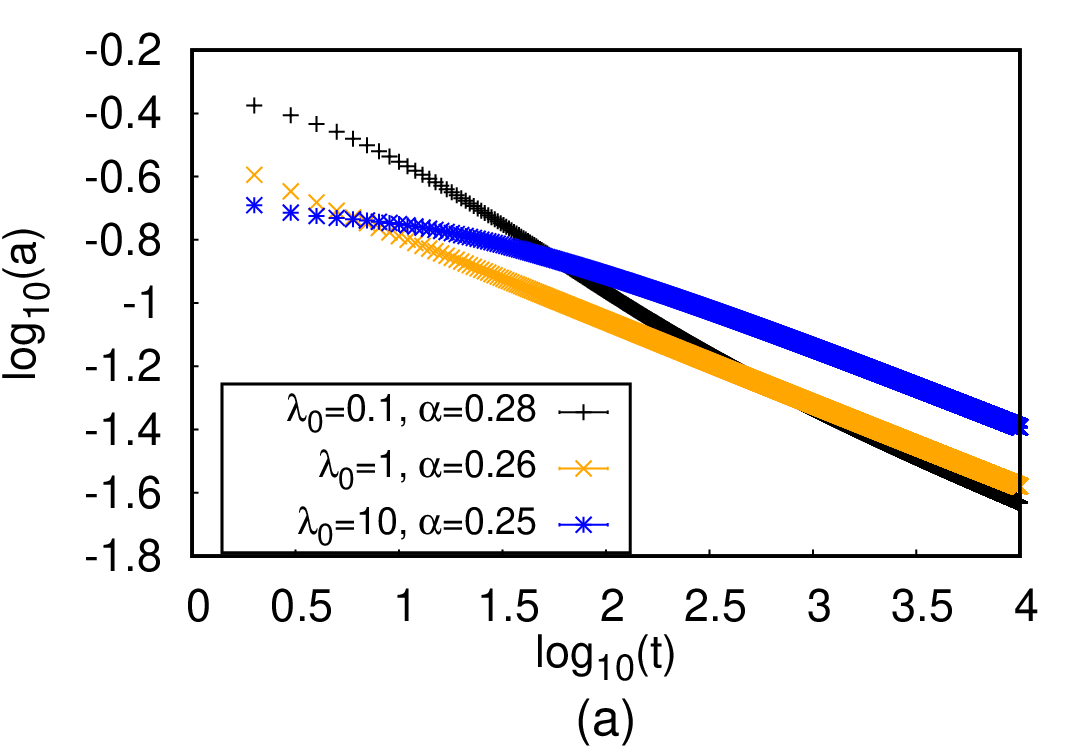} \
    \includegraphics[width=0.66\columnwidth]{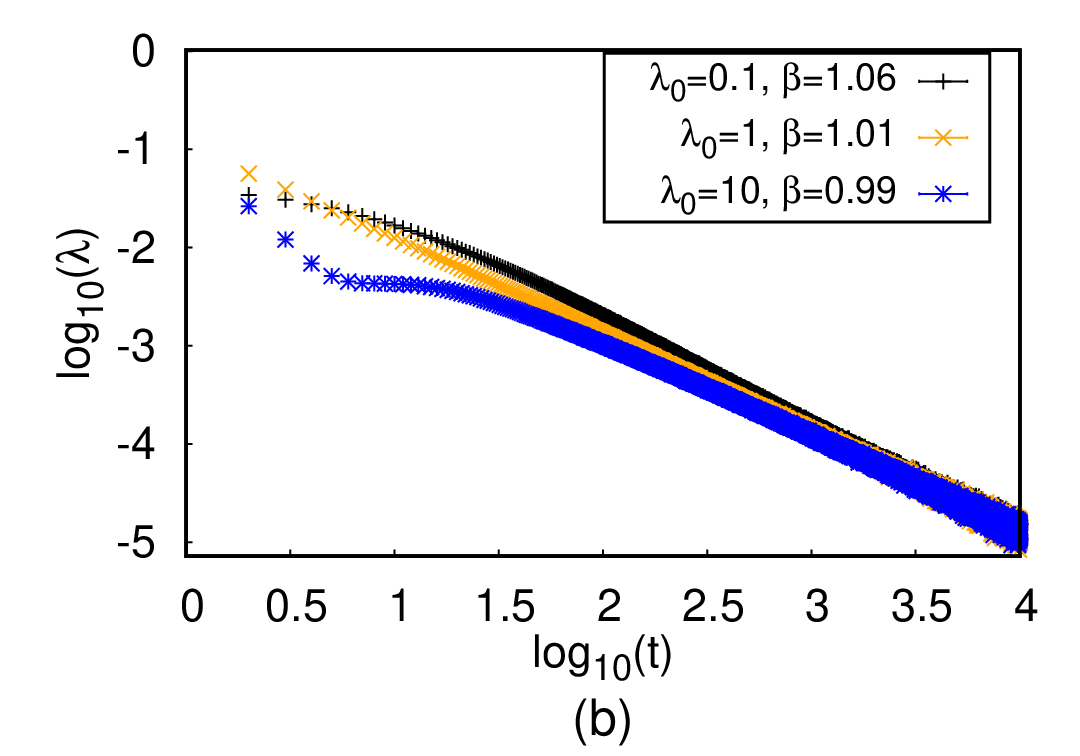} \
    \includegraphics[width=0.66\columnwidth]{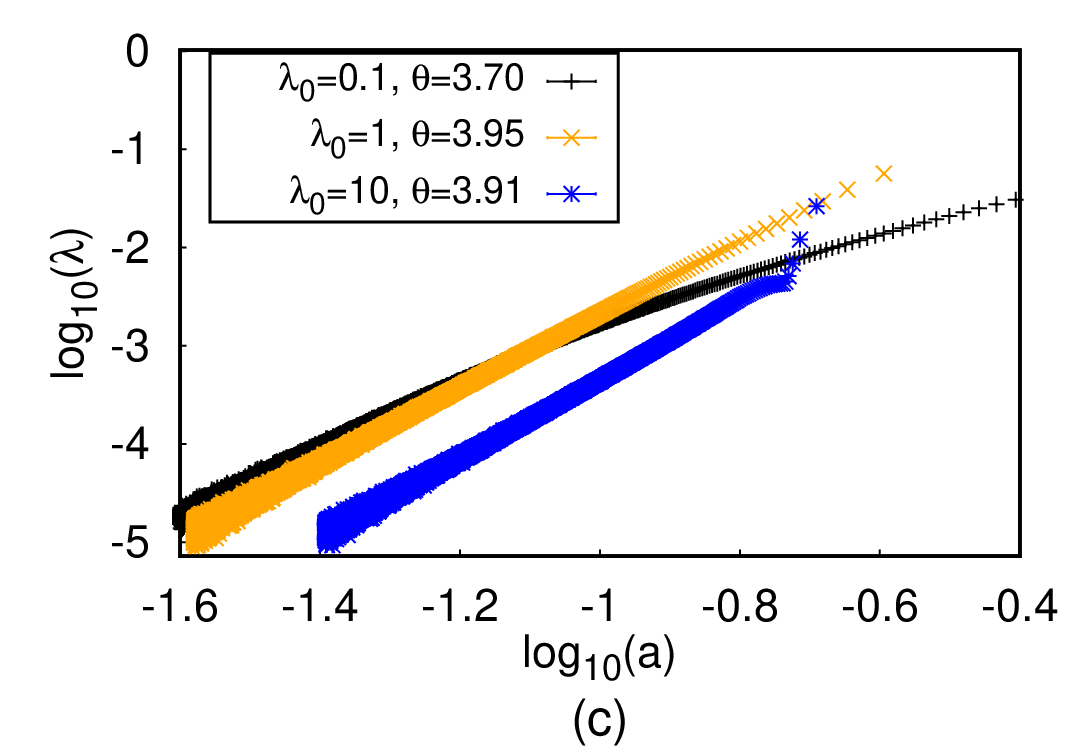}
    \caption{Two-species pair annihilation kinetics $A + B \xrightarrow{\lambda} \emptyset$ on a 
	one-dimensional lattice of size $L = 100,000$ with equal initial densities $a(0) = b(0) = 1$, for 
	different microscopic reaction rates $\lambda_0 = 0.1$, $1$, and $10$, as indicated: 
	Double-logarithmic plots of (a) the particle density decay $a(t)$; the macroscopic effective reaction 
	rate as function of (b) time $\lambda(t)$, and (c) density $\lambda(a)$.
	The resulting effective scaling exponents $\alpha$, $\beta$, and $\theta$ are listed. 
	The data were averaged over $1000$ independent simulations.}
\label{fig:ann_1d}
\end{figure*}
In summary, our Monte Carlo simulations for diffusive pair coagulation processes in $d = 1$, $2$, and 
$3$ dimensions demonstrate that our algorithm successfully generates the correct long-time scaling 
properties for the macroscopic reaction rate $\lambda$, but in addition allows the investigation of 
nonuniversal crossover features before the asymptotic regimes are reached. 
This indicates that counting the number of events, normalizing properly, and collecting adequate statistics 
provides an accurate means to estimate reaction rate renormalization effects for physical, (bio-)chemical, 
ecological, and epidemiological systems subject to nonlinear stochastic reactive processes.

\subsection{\label{section:TwoSpeciesAnnihilationResults} Two-species pair annihilation}

Two-species (e.g., particle-antiparticle) pair annihilation is commonly described as a reaction-diffusion 
system with the single binary reaction  $A + B \xrightarrow{\lambda} \emptyset$. 
Since $A$ and $B$ particles are simultaneously removed from the system during any annihilation reaction,
their particle number difference $c = a(t) - b(t)$ is a locally conserved quantity, implying that one only 
needs to study the behavior of, say, $b(t)$, and the conservation law can be used to directly infer $a(t)$. 
The system exhibits different behavior depending on whether $c = 0$ or $c \neq 0$. 
For unequal initial species numbers $c > 0$, the minority population density decays according to 
$\ln{b(t)} \sim - t^\alpha$, where $\alpha = d / 2$ yielding a stretched exponential for $d < d_c = 2$,
while $\alpha = 1$ above the critical dimension, and with logarithmic corrections appearing at $d_c = 2$.
The majority species approaches its asymptotic density in a similar manner, 
$\ln{a(t) - c} \sim - t^\alpha$.
Yet in this work, we focus on the special situation $c = a(t) - b(t) = 0$; then the (stretched) exponential 
behavior is replaced by power laws $a(t) = b(t) \sim t^{-\alpha}$ \cite{Lee1, Lee3, Lee4, CogAnn4, 
Ann1, Ann2}. 
In this symmetric ``critical'' situation, the system is governed by a single macroscopic rate equation 
(\ref{eq:coagre}) for both $A$ and $B$ particle densities.

The macroscopic rate is calculated by counting the number of reactions occurring at each time step divided 
by the total particle number $N_A + N_B$ then present, since the annihilation process may be initiated by  either an $A$ or $B$ individual in the simulation. 
This prescription of determining the macroscopic rate $\lambda$ is consistent with the rate equation 
(\ref{eq:coagre}). 
Again we implement an off-site algorithm here, to avoid artifacts at initially high densities.

As the rate equation for $A + B \to \emptyset$ is identical with the one for $2 A \to A$, the ensuing
long-time scaling may be described by the same exponents $\alpha$, $\beta = 1$, and 
$\theta = 1 / \alpha$. 
However, these exponets assume different values for single-species pair coagulation and two-species 
annihilation.
While the renormalizations describing depletion zones still apply in dimensions $d \leq d_c = 2$, the two 
distinct species may now segregate into two separate $A$ and $B$ domains, wherein no reactions take 
place. 
Annihilation processes are thus confined to the boundaries of the $A$- and $B$-rich zones, which leads to 
a further drastic slowing-down of the decay dynamics in dimensions $d < d_s = 4$ \cite{CogAnn2, 
CogAnn6, uwesbook}.
The resulting density decay exponents induced by species segregation are $\alpha = d / 4$ and 
$\theta = 4 / d$.  
For $d > d_s = 4$, the mean-field exponents $\alpha = \beta = \theta = 1$ apply.
It should be noted that species segregation is an asymptotic effect that can only be observed at sufficiently
long times; at intermediate time regimes, the system may show the diffusion-limited depletion scaling.

For two-species pair annihilation, here we only focus on the one-dimensional case, since the scaling 
regime is more difficult to access in higher dimensions, and our principal aim in this section is to 
demonstrate the viability of our technique for reactive dynamics involving multiple species. 
Figure~\ref{fig:ann_1d} demonstrates that our simulation data reproduce the expected scaling behavior
for $\lambda_0 = 1$ or bigger.
For small $\lambda_0 = 0.1$, we find effective exponents that are still in the crossover interval between 
the diffusion-limited depletion and segregation values.

\section{\label{section:PredatorPreyResults} Lotka--Volterra predator-prey competition and coexistence}

In the previous sections, we have demonstrated that extracting the statistics for the number of reactions in 
a Monte Carlo simulation produces the expected behavior for the coarse-grained macroscopic reaction 
rates. 
Next we consider a system where in contrast little is known about the relationship between the microscopic 
and macroscopic parameters. 
The paradigmatic Lotka--Volterra model for predator-prey competition and coexistence comprises two 
species: the predators $A$ and prey $B$. 
Left alone, the predator population is subject to spontaneous death processes 
$A \xrightarrow{\mu} \emptyset$, and hence would decay exponentially over time with the rate $\mu$.
By themselves, the prey reproduce according to the linear branching or birth reaction 
$B \xrightarrow{\sigma} 2 B$ that would cause a Malthusian population explosion with rate $\sigma$.
The nonlinear binary predation reaction $A + B \xrightarrow{\lambda} 2 A$, whereupon predators 
immediately reproduce after a successful ``hunt'', replacing a prey individual with a predator offspring,
controls the prey density and opens the possibility of coexistence for both species \cite{lotka, volterra, 
uwesbook, intro6, intro8, intro11, intro13, intro14, intro15, intro16, exponents1, exponents2, 
exponents3, exponents4, exponents5, ftexponents1, LV1, LV2, LV3, LVme}.
In order to prevent an exponential divergence of the prey density in the absence of predators, in our lattice
simulations we prescribe an on-site restriction on the total particle number $N_A(x,t) + N_B(x,t) \leq K$, 
which globally translates to a finite carrying capacity for the total population $a(t) + b(t) \leq K$ that holds 
also for the local mean species densities~\footnote{We could instead only restrict the local prey density 
$b(x,t) \leq K$; yet this modification does not significantly change the system's behavior.}.
Correspondingly, in mean-field approaches the prey population constraint is typically represented through 
a logistic term in their rate equation. 
Yet as we have shown in Sec.~\ref{section:BirthResults}, this is adequate only in the long-time limit. 
The stochastic processes in our lattice Monte Carlo simulations consist of the above death, birth, and 
predation reactions, in addition to nearest-neighbor hopping for both species. 
We employ the off-site implementation for reproduction and predation processes. 
Hence the prey offspring production happens on one of the neighboring sites, and the predation reaction 
occurs if a predator finds a prey on an adjacent site~\footnote{If a prey is selected and finds a predator 
on a neighboring site, a predation reaction is not attempted.}, which effectively induces hopping transport 
via proliferation \cite{intro13,ftexponents1,LV3,LVme}.

Furthermore, we let a predator attempt a predation reaction for each prey individual located on the chosen 
neighboring site. 
Consequently, we compute the macroscopic prey birth rate by dividing the number of branching events in
the MCS by the total prey number $N_B$ at that instant, and the macroscopic predator decay rate by 
dividing the number of death events by the total predator number $N_A$.
For the binary predation reaction, we instead calculate the more appropriate effective coarse-grained 
``coupling'' $\lambda$ as the ratio of the number of predation events in each MCS and the product 
$N_A \, N_B$.
The above algorithm motivates us to write the associated macroscopic reaction-diffusion equations for the 
local predator and prey densities $a(x,t)$, $b(x,t)$ in the Lotka--Volterra model in the form
\begin{subequations}
\label{eqs:LV}
\begin{align}
	&\frac{\partial a}{\partial t} = D_a(a,b) \, \nabla^2 a + \lambda(a,b) \, a \, b - \mu(a,b) \, a \ , \\
	&\frac{\partial b}{\partial t} = D_b(a,b) \, \nabla^2 b + \sigma(a,b) \, b - \lambda(a,b) \, a \, b \ .
\end{align}
\end{subequations}

\begin{figure*}[t]
    	\includegraphics[width=0.66\columnwidth]{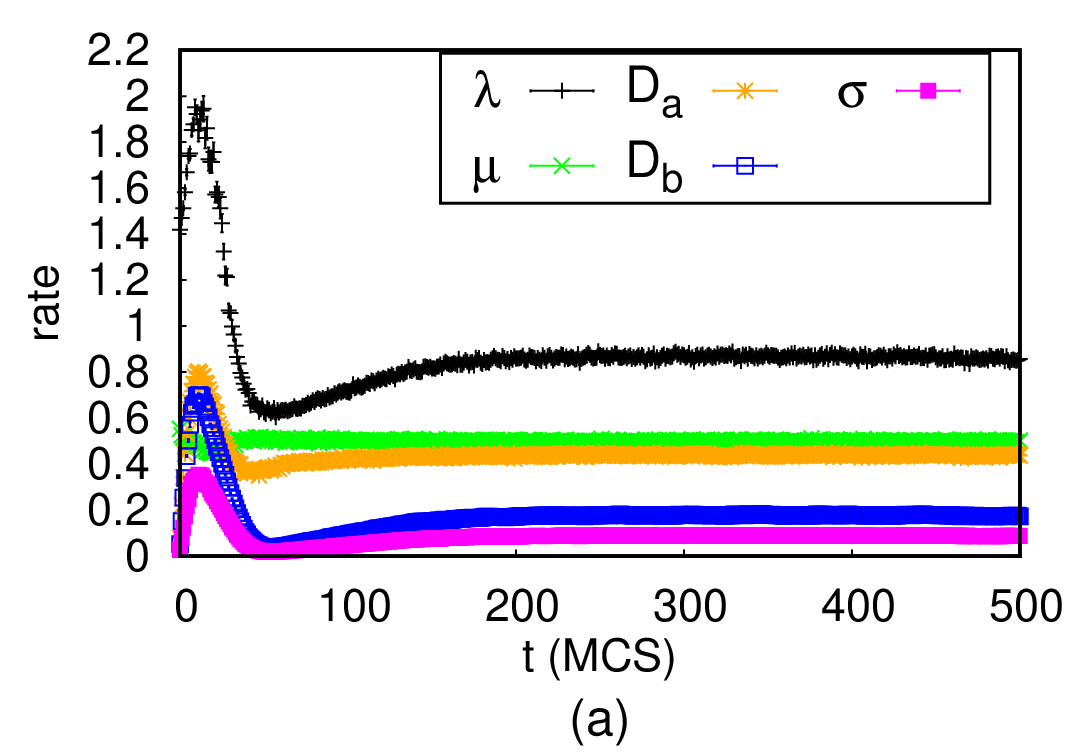} \
    	\includegraphics[width=0.66\columnwidth]{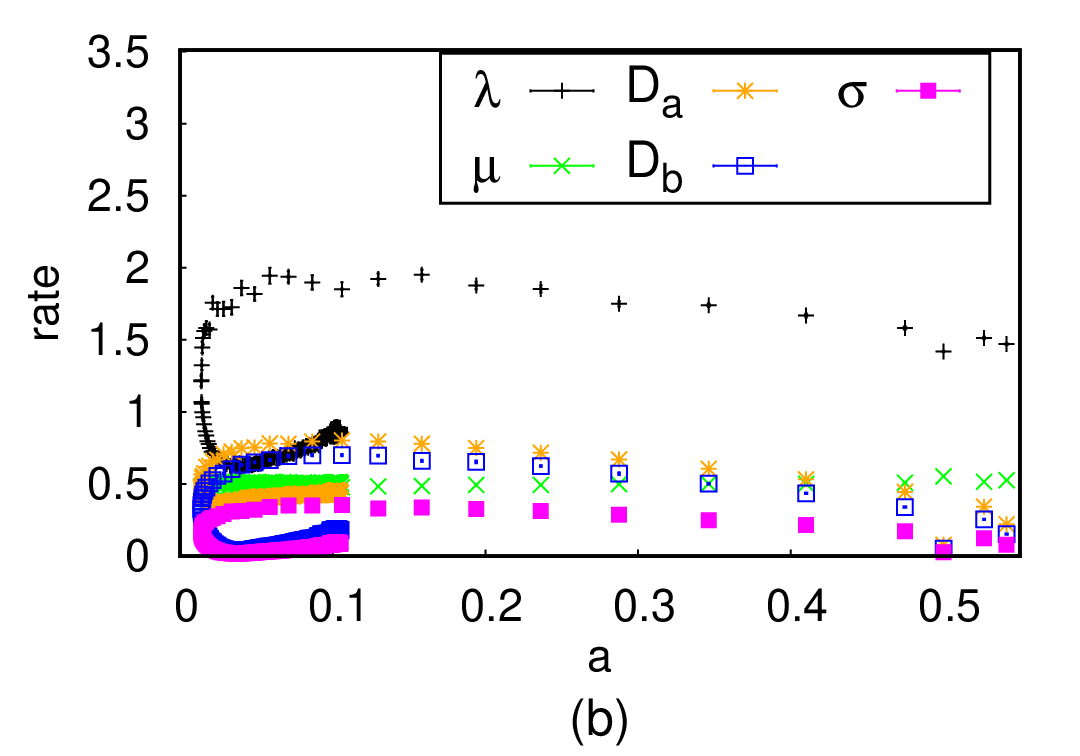} \
    	\includegraphics[width=0.66\columnwidth]{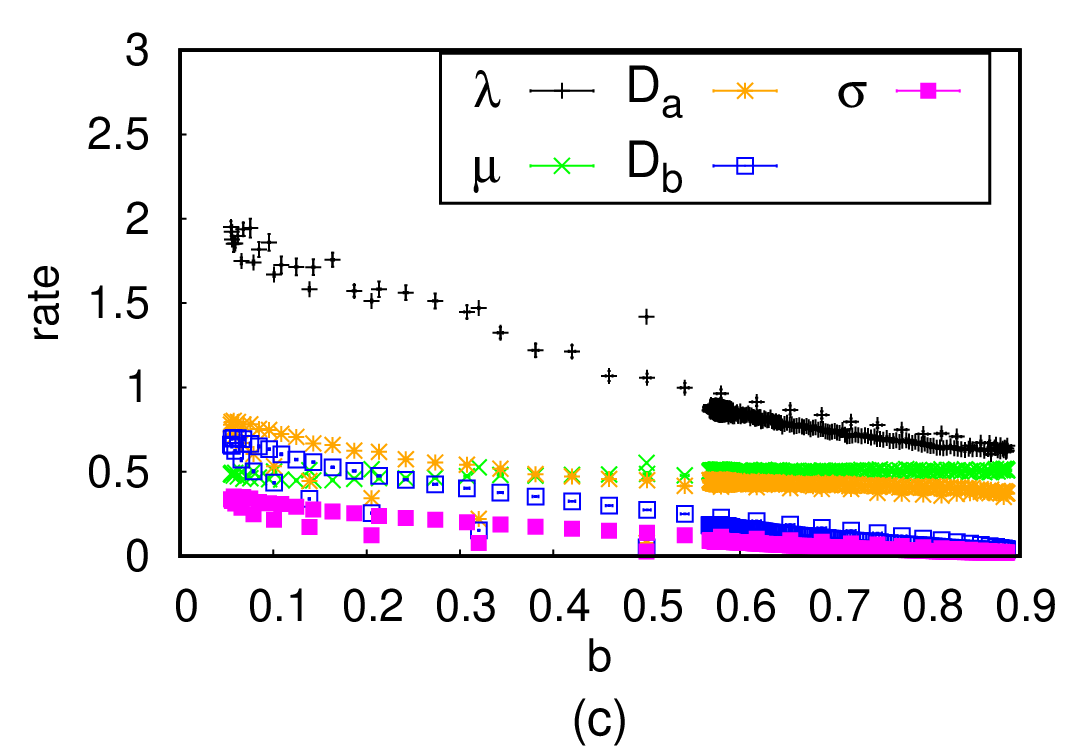} 
\caption{Macroscopic reaction rates $\mu$, $\sigma$, coupling $\lambda$, as well as diffusivities $D_a$, 
	$D_b$ for the stochastic Lotka--Volterra predator-prey model on a two-dimensional square lattice 
	(with periodic boundary conditions) with length $L = 150$, initial population densities 
	$a(0) = b(0) = 0.5$, microscopic propensities $\mu_0 = \sigma_0 = 0.5$, $\lambda_0 = 0.8$, 
	$D_0 = 1$, and carrying capacity (on-site occupation restriction) $K = 1$, displayed as functions of 
	(a) time $t$ (in MCS); (b) predator density $a$; (c) prey density $b$.
	The data were averaged over $20$ independent Monte Carlo simulation runs.}
\label{fig:LV_r}
\end{figure*}
In the mean-field rate equation approximation $\lambda$ and $\mu$ are taken to be constants, whereas 
$\sigma(a,b) = \sigma_0 \left( 1 - \left(a+b\right) / K \right)$. 
The resulting coupled system of ordinary differential equations has three stationary 
solutions (fixed points): 
(i) Total extinction, where both predator and prey populations decay to zero.
This configuration is always unstable in the mean-field approximation.
(ii) Predator extinction and thus prey fixation, where the predator population disappears, and the prey 
fill out the entire lattice (up to the carrying capacity $K$). 
Note that in any finite stochastic system, either of these absorbing configurations is ultimately attained,
namely (i) or (ii) if either the prey or predators happen to go extinct first. 
(iii) Species coexistence, where both species ultimately survive at non-zero density values. 
If a finite (prey or total) carrying capacity $K > 0$ is implemented, there appears an active-to-absorbing 
phase transition that occurs at some critical predation rate $\lambda_c$: 
If the predators are not able to predate quickly enough, they will die out and the system goes into prey 
fixation (ii). 
Above this critical $\lambda_c$ value, the more efficient predators are able to reproduce sufficiently fast
to coexist alongside the prey. 
The resulting mean-field dynamical system also exhibits nonlinear population oscillations as there 
emerges a periodic cycle of growing number of prey, followed by an increase in the predator population, 
leading to a decrease in the number of prey, and in consequence of the predators as well as their food
resources become sparse, and then the cycle repeats as for low predator prevalence, the prey reproduce
fast again. 
In this coexistence phase, the spatially extended mean-field reaction-diffusion equations allow for 
propagating wave solutions of the form $a(x,t) = a(x-vt)$, $b(x,t) = b(x-vt)$ with wavefront speed $v$. 
These waves are induced by the prey attempting to evade the predators by effectively reproducing at a 
safe distance from them, and the predators pursuing their prey as they can only reproduce effectively in
regions of high prey concentration.
However, mean-field theory predicts that those wave solutions are unstable and decay, leaving the
homgeneous population distribution as the only stable solution.

\begin{figure}[b]
    	\includegraphics[width=0.9\columnwidth]{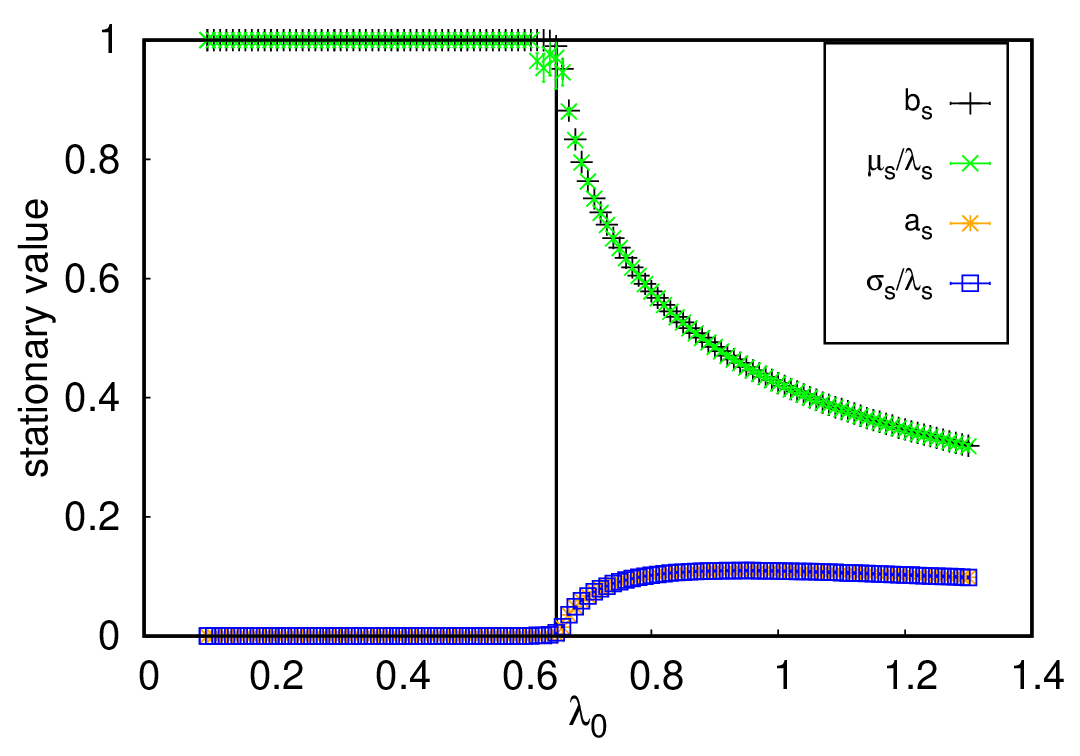} 
\caption{Long-time stationary population densities $a_s$, $b_s$ and corresponding macroscopic rate 
	ratios $\sigma_s / \lambda_s$, $\mu_s / \lambda_s$ for the stochastic Lotka--Volterra predator-prey 
	model on a square lattice with length $L = 150$, initial densities $a(0) = b(0) = 0.5$, and 
	microscopic rates set to $\mu_0 = \sigma_0 = 0.5$, $D_0 = 1$, with carrying capacity $K = 1$, 
	plotted as a function of the microscopic predation rate $\lambda_0$. 
	The black vertical line indicates the location of the active-to-absorbing state transition that constitutes 
	a predator extinction / prey fixation threshold.
	The data were averaged over $20$ independent simulation runs.}
\label{fig:LV_fp_comparison}
\end{figure}
This qualitative picture remains intact in the stochastic lattice system, yet with drastic 
quantitative deviations \cite{intro6, intro8, intro9, intro11, intro12, intro13, intro14, intro15, intro16, 
exponents1, exponents2, exponents3, exponents4, exponents5, ftexponents1, LV1, LV2, LV3, LVme}. 
In a non-spatial setting, intrinsic multiplicative reaction and demographic noise in the two-species
coexistence state effectively maps to additive white noise driving the nonlinear population oscillations
through a resonant amplification mechanism \cite{mf3}. 
In spatially extended systems, the inevitable dynamical correlations generated by the stochastic 
reaction processes induce a dependence of the macroscopic rates on the population densities. 
Indeed, the Lotka--Volterra predator-prey model exhibits pronounced spatial correlations that are 
manifest in persistent pursuit and evasion waves that originate from randomly distributed prey 
survivors following sweeps of predation events:
The stochastic ``kicks'' in the system are now sufficiently strong to render the wave solutions stable 
in the long-time limit.
In addition, and in accord with general expectations, the continuous active-to-absorbing phase 
transition is found to be governed by the directed percolation universality class \cite{intro13, intro14, 
exponents1, exponents2, exponents3, exponents4, exponents5, ftexponents1, LV2}, with universal
critical exponents that deviate from the mean-field values in dimensions $d < d_c = 4$.
For a more thorough review of stochastic spatially extended Lotka--Volterra systems, we refer to the
overview in Ref.~\cite{intro13}.

\begin{figure*}[t]
    	\includegraphics[width=0.85\columnwidth]{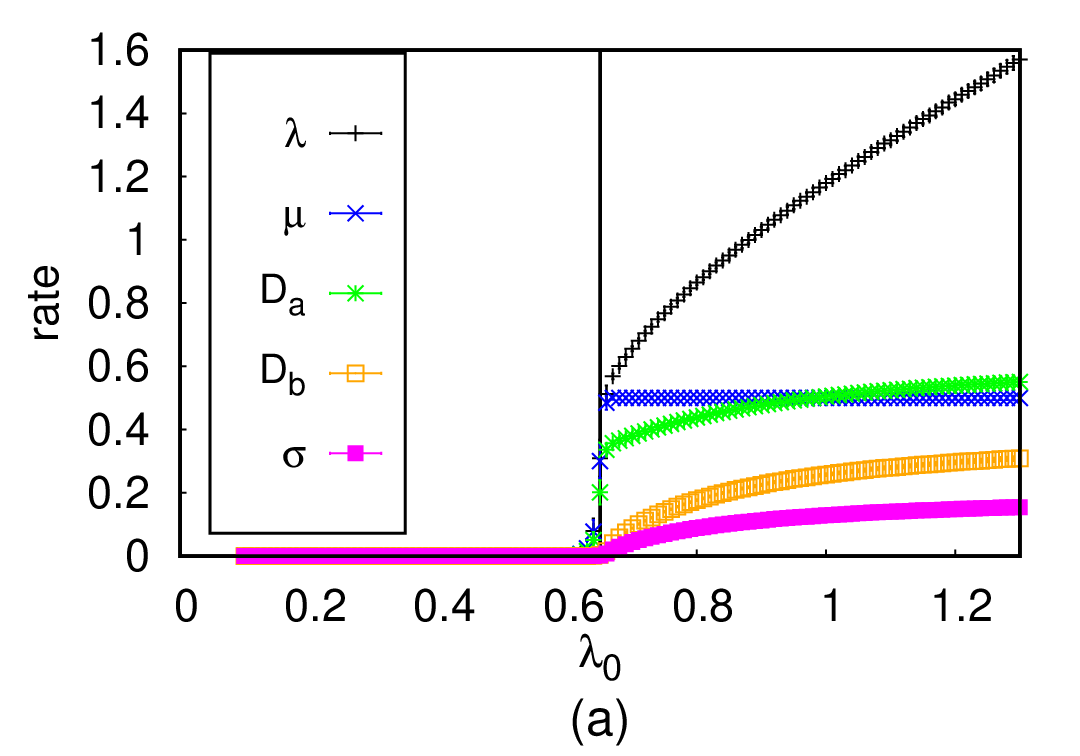} \
    	\includegraphics[width=0.85\columnwidth]{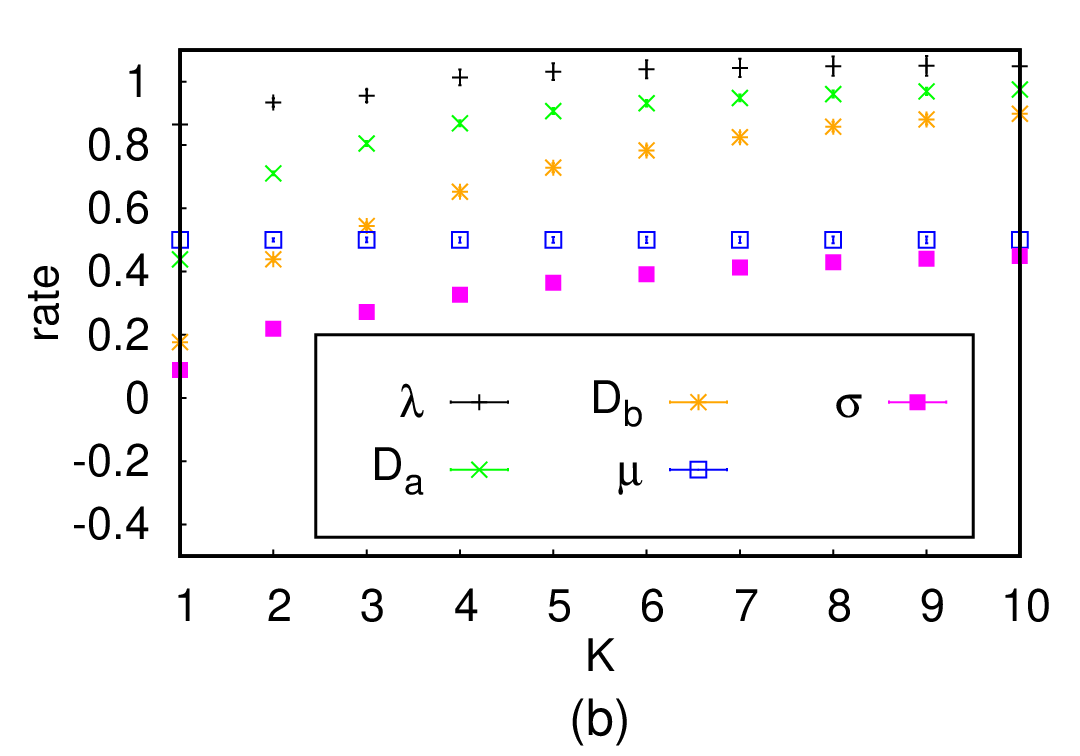} \\
    	\includegraphics[width=0.85\columnwidth]{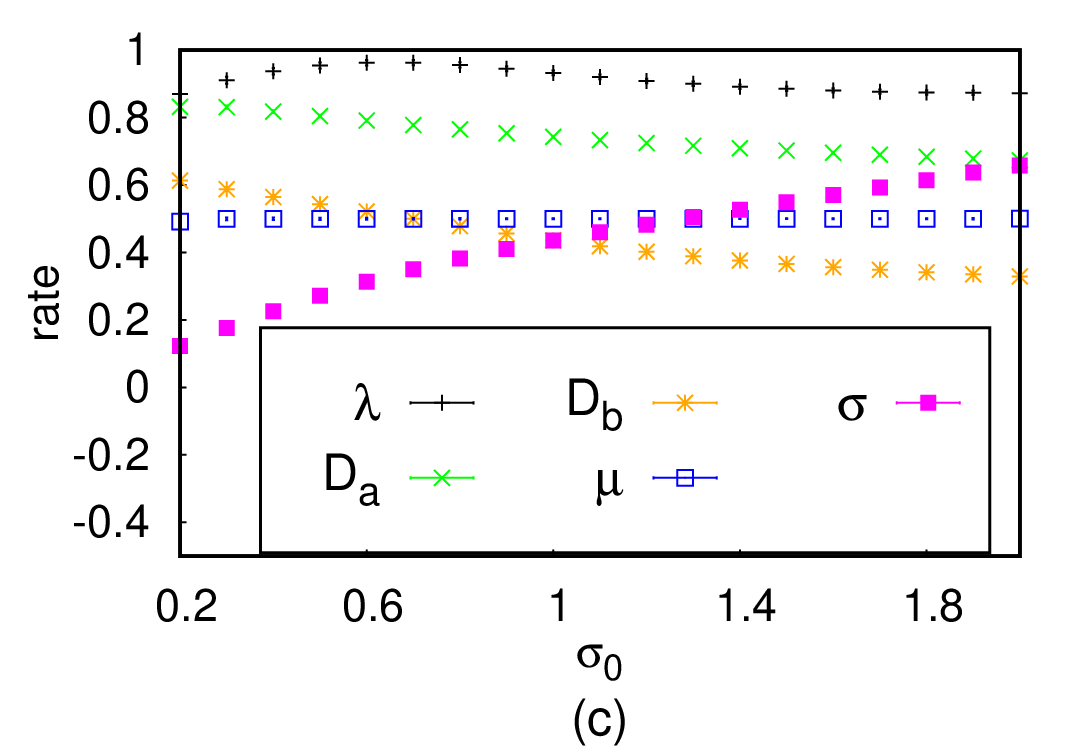} \
    	\includegraphics[width=0.85\columnwidth]{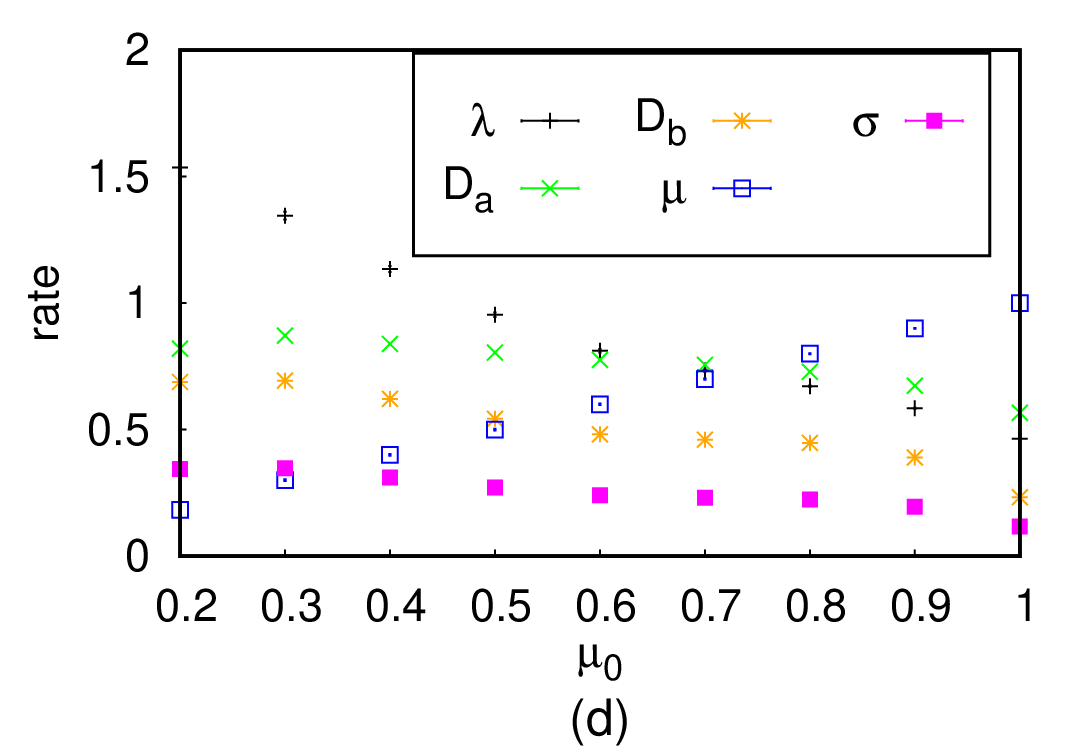}
\caption{Macroscopic reaction rates $\mu$, $\sigma$, coupling $\lambda$, as well as diffusivities $D_a$, 
	$D_b$ for the stochastic Lotka--Volterra predator-prey model on a square lattice with length 
	$L = 150$, initial densities $a(0) = b(0) = 0.5$, as functions of the microscopic input parameters: 
	Dependence (a) on $\lambda_0$, with fixed $\mu_0 = \sigma_0 = 0.5$, $K = 1$; (b) on $K$, with 
	$\lambda_0 = 0.8$, $\mu_0 = \sigma_0 = 0.5$; (c) on $\sigma_0$, with $\lambda_0 = 0.8$, 
	$\mu_0 = 0.5$, $K = 3$; (d) on $\mu_0$, with $\lambda_0 = 0.8$, $\sigma_0 = 0.5$, $K = 3$. 
	The vertical black line in figure (a) marks the predator extinction / prey fixation threshold.
	The data were averaged over $20$ independent simulation runs.}
\label{fig:LV_r_vs_pars}
\end{figure*}
In this model, the sole reaction that can occur without any restrictive conditions is predator death. 
Hence its behavior can be predicted \textit{a priori} as follows: 
The expected number of death reactions occurring in a single MCS is 
$\langle \text{number\,of death events} \rangle = \left( N_A + N_B \right) p(\text{pick predator})$ 
$p(\text{death reaction}\,|\,\text{predator is picked})$. 
But the (average) probabilities for selecting a predator and then picking a death reaction are given by their 
relative propensities, which are easily obtained:
\begin{align*}
	&\, p(\text{pick\,predator}) = \frac{N_A \left( D_0 + \mu_0 + \lambda_0 \right)}
	{N_A \left( D_0 + \mu_0 + \lambda_0 \right) + N_B \left( D_0 + \sigma_0 \right)} , \\
	&\ p(\text{death\,reaction}\,|\,\text{predator\,is\,picked}) 
	= \frac{\mu_0}{D_0 + \mu_0 + \lambda_0} \ .
\end{align*}
As mentioned in Sec.~\ref{section:Methods}, we set $D_0 = 1$; thus we obtain
\begin{equation}
\label{eqs:timeScaling}
	\mu = \frac{\left( N_A + N_B \right) \mu_0}
	{N_A \left( 1 + \mu_0 + \lambda_0 \right) + N_B \left( 1 + \sigma_0 \right)} \ .
\end{equation}
Here we may of course replace the particle numbers $N_A$ and $N_B$ by the predator and prey densities 
$a$ and $b$.
The macroscopic reaction rate $\mu$ is then fully determined by Eq.~(\ref{eqs:timeScaling}). 
Intuitively, the death reaction constitutes a linear stochastic process, and therefore its macroscopic rate 
should not depend on the other microscopic propensities $\lambda_0$ and $\sigma_0$. 
However, due to our chosen normalization with all propensitites adding up to unity, the linear death
processes pick up dependences on the other rates; this could be avoided by choosing a different algorithm
or parameterization. 
Regardless, employing normalized rates that sum to unity is computationally convenient. 
One may interpret the denominator in Eq.~(\ref{eqs:timeScaling}) as a multiplicative time step rescaling 
factor that could be absorbed into all extracted rates. 
Consequently, calculating the mean macroscopic death rate $\mu$ after this rescaling, we expect that it 
should match precisely with its microscopic value $\mu_0$. 
Henceforth, all reported macroscopic rate values are divided by this factor, unless otherwise specified.

Figure~\ref{fig:LV_r} shows how the effective macroscopic rates $\mu$, $\sigma$, coupling $\lambda$,
and diffusivities $D_a$, $D_b$ vary with time $t$ and the predator and prey densities $a$, $b$ in our
Monte Carlo simulations on a two-dimensional lattice (with periodic boundary conditions). 
The temporal evolution of the coarse-grained parameters tracks the dynamics one would observe for the 
population densities, consistent with the macroscopic rates being density-dependent functions. 
In the species coexistence regime, the macroscopic parameters all settle down to stationary values, as seen
in Figure~\ref{fig:LV_r}(a). 
The macroscopic predator death rate $\mu$ displays no noticeable dependence on either population 
density, remaining constant in time and equal to its microscopic value $\mu_0$. 
Note that even though the rates appear to be multi-valued as a functions of the predator and prey densities 
separately in Figs.~\ref{fig:LV_r}(b,c), this is an artifact of them actually being functions of both $a$ and 
$b$: 
Plotting the macroscopic parameters in a three-dimensional plot as functions of $a$ and $b$ reveals their
single-valuedness. 
Although the coarse-grained rates do not represent the system's degrees of freedom, these graphs of the 
macroscopic parameters as functions of the population densities are reminiscent of phase space plots.

The stationary values of the macroscopic rates $\mu_s$, $\sigma_s$ and coupling $\lambda_s$ can be 
used to predict the densities' ``fixed-point'' values $a_s$ and $b_s$. 
Examining eqs.~(\ref{eqs:LV}) and setting the time derivatives to zero we obtain, akin to simple 
mean-field theory, $a_s = \sigma_s / \lambda_s$ and $b_s = \mu_s / \lambda_s$. 
Recall that the macroscopic prey birth rate incorporates the carrying capacity restriction through its 
nontrivial density dependence, similar to the restricted birth model of Sec.~\ref{section:BirthResults}; 
consequently, $\sigma$ implicitly becomes a function of $K$. 
This is confirmed in Fig.~\ref{fig:LV_fp_comparison}, where the stationary population densities are 
determined directly from the simulations and then compared against the numerically computed rate ratios 
$\sigma_s / \lambda_s$ and $\mu_s / \lambda_s$. 
Intriguingly, this comparison yields almost perfect agreement between the rate equation formulas and the 
actual stationary density values.
We discern marked deviations of the actual value for $b_s$ from the ratio $\mu_s / \lambda_s$ only in 
the vicinity of the predator extinction threshold: 
Stochastic fluctuations and critical correlations become prominent near the active-to-absorbing phase 
transition, and their effects are not fully accounted for by mere renormalizations of the effective 
parameters.

Extracting the macroscopic rates and coupling over the course of the Monte Carlo simulations allows us to 
investigate in a straightforward manner their dependence on the microscopic input parameters 
$\lambda_0$, $K$, $\sigma_0$, and $\mu_0$, as depicted in Fig.~\ref{fig:LV_r_vs_pars}.
Figure~\ref{fig:LV_r_vs_pars}(a) shows that $\sigma$ and $D_b$ are continuous at the predator 
extinction / prey fixation critical point, while $\lambda$, $\mu$, and $D_a$ exhibit a jump discontinuity
there. 
In the absorbing state where the lattice is completely filled with prey, all reactions cease.
As anti\-cipated, the macroscopic death rate $\mu$ is not altered from its microscopic value $\mu_0$.
The macroscopic prey birth and hopping rates $\sigma$ and $D_b$ become diminished as the absorbing 
state transition is approached, and the lattice increasingly saturated with the $B$ species, much like in the
restricted birth model of Sec.~\ref{section:BirthResults}; indeed, the ratio $\sigma / D_b$ remains 
constant.
Generally, our simulation results yield mostly monotonic dependences of the macroscopic rates on the 
microscopic parameters. 
The exception is the functional dependence of $\lambda$ on $\sigma_0$, shown in 
Fig.~\ref{fig:LV_r_vs_pars}(c), which displays a maximum. 
This feature can be explained by noting that as $\sigma_0$ is raised, the prey number increases, in turn
causing an enhancement in $\lambda$; yet past a certain value, the predators quickly consume most of 
the prey, such that once the system reaches its stationary state, fewer prey remain available, leading to a 
decrease of the effective predation coupling $\lambda$. 
As seen in Figs.~\ref{fig:LV_r_vs_pars}(a,b) an increase in either $\lambda_0$ or $K$ induces faster 
reactions. 
In contrast, raising $\sigma_0$ causes a decrease in the renormalized diffusivities $D_a$ and $D_b$, 
shown in Fig.~\ref{fig:LV_r_vs_pars}(c), as there are fewer lattice vacancies available for hopping. 
Similarly, as depicted in Fig.~\ref{fig:LV_r_vs_pars}(d), increasing $\mu_0$ slows down diffusion, and 
also reduces the branching rate $\sigma$, since the prey population is more likely to reach saturation as 
the predators die away fast.

\begin{figure}[t]   
    	\includegraphics[width=0.95\columnwidth]{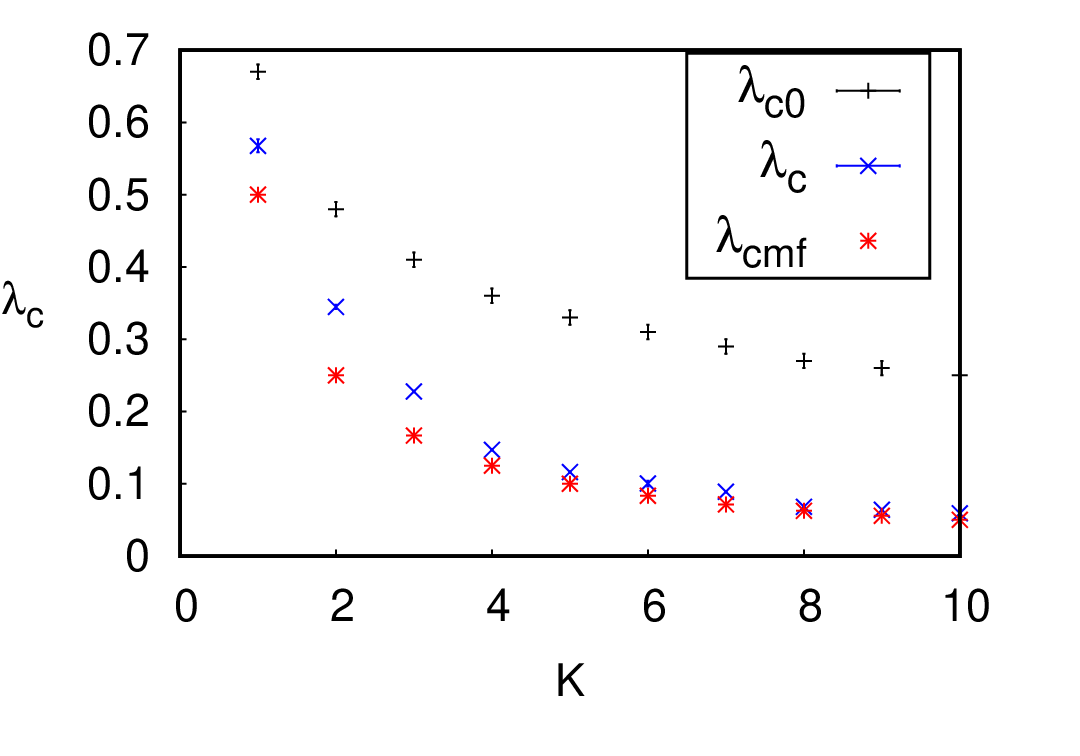} 
\caption{Predator extinction / prey fixation threshold for the stochastic Lotka--Volterra predator-prey 
	model on a square lattice with length $L = 150$ and initial densities $a(0) = b(0) = 0.5$ as function 
	of the local occupation restriction (carrying capacity) $K$, for fixed microscopic rates 
	$\mu_0 = \sigma_0 = 0.5$. 
	Here, $\lambda_{c0}$ denotes the microscopic threshold; 
	$\lambda_\text{cmf} = \mu / K = \mu_0 / K$ is the threshold value predicted by the mean-field 
	rate equations; and $\lambda_c$ represents the corresponding macroscopic value obtained from the
	Monte Carlo simulations.
	The data were averaged over $20$ independent simulation runs.}
\label{fig:LV_l_c}
\end{figure}
We can also utilize our technique to test the mean-field formula for the location of the critical 
active-to-absorbing transition point. 
For fixed values of $\mu$ and $\sigma$, the rate equations predict a predator extinction and prey fixation
threshold at $\lambda_\text{cmf} = \mu / K$. 
In Fig.~\ref{fig:LV_l_c}, we compare the ``microscopic'' transition point $\lambda_{c0}$, given by the 
microscopic predation probability used to perform the Monte Carlo simulations, with the numerically 
determined actual threshold location $\lambda_c$ obtained from our event statistics in the Monte Carlo 
simulation data, and additionally the value $\lambda_\text{cmf}$ predicted by the mean-field rate 
equations with the renormalized macroscopic rates, for a series of different carrying capacities 
$K = 1, \ldots, 10$.
We find that $\lambda_\mathrm{cmf}$ is closer to the actual transition point $\lambda_c$ than 
$\lambda_{c0}$.
Interestingly, these curves can be approximately fit with power laws:
The microscopic transition scales according to $\lambda_{c0} \sim K^{- 0.43}$,
whereas the numerically determined ``macroscopic'' predator extinction threshold behaves as 
$\lambda_c \sim K^{- 0.94}$, closer to the mean-field rate equation dependence $\sim K^{-1}$. 
This indicates that the actual control parameter determining the critical point is more adequately captured
by the macroscopic predation rate $\lambda$ than the microscopic probability $\lambda_0$ implemented 
in the Monte Carlo simulation algorithm. 

\begin{figure}[t]
    	\includegraphics[width=0.95\columnwidth]{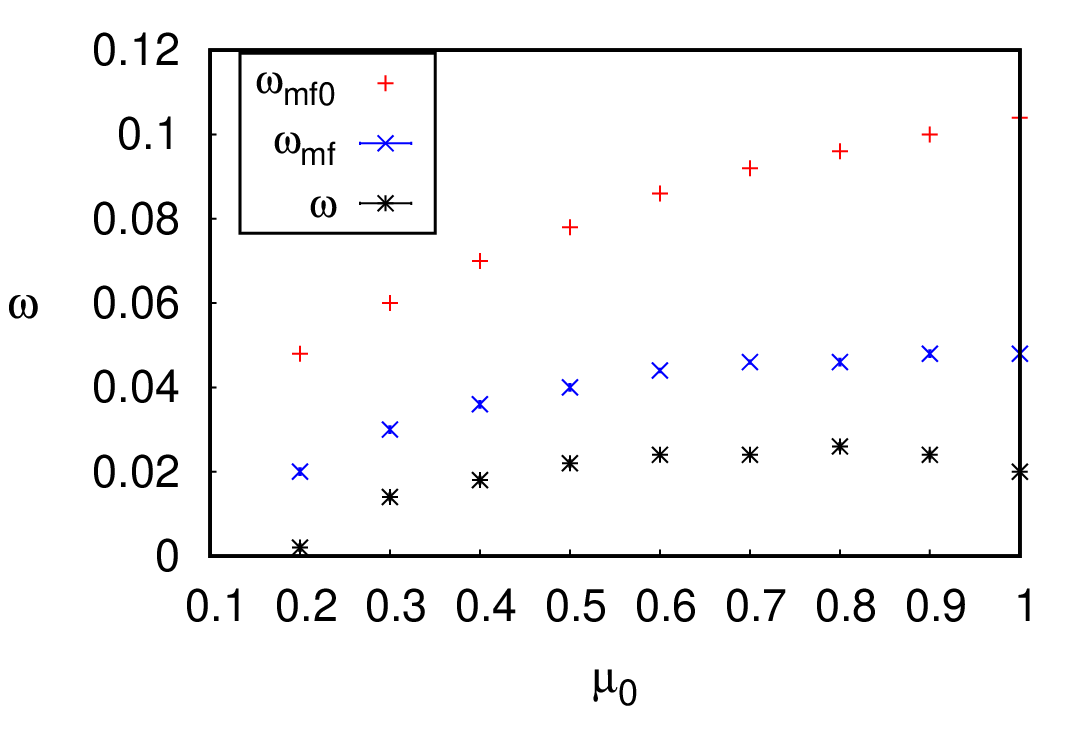} 
\caption{Characteristic population oscillation frequency $\omega$ for the stochastic Lotka--Volterra 
	predator-prey model on a square lattice with length $L = 150$ and initial densities 
	$a(0) = b(0) = 0.5$ as function of the predator death rate $\mu_0$, with fixed parameters 
	$\sigma_0 = 0.5$, $\lambda_0 = 0.8$, and $ K = 10$. 
	For comparison, the oscillation frequencies $\omega_\text{mf0}$ and 
	$\omega_\text{mf}$ as determined by numerically solving the nonlinear coupled rate equations 
	(\ref{eqs:LV}) using respectively the microscopic and macroscopic rate parameters are depicted as 
	well.
	The data were averaged over $10$ independent simulation runs.}
\label{fig:LV_freq}
\end{figure}
It is well-established that the population oscillation frequency in the two-species coexistence phase of the 
Lotka--Volterra predator-prey system exhibits strong fluctuation corrections relative to its mean-field value 
\cite{intro12,intro13,LV2}.
Using our Monte Carlo simulations, we have determined the actual oscillation frequency $\omega$ from 
the peak location in the Fourier transform of the population density time tracks.
In Fig.~\ref{fig:LV_freq} this numerically extracted frequency is compared against the ``microscopic'' and 
``macroscopic'' rate equation oscillation frequencies $\omega_\text{mf0}$ and $\omega_\text{mf}$,
respectively, graphed as functions of the predator death rate $\mu_0$. 
Here, $\omega_\text{mf0}$ was computed through integrating the rate equations (\ref{eqs:LV}) by 
means of a fourth-order Runge--Kutta scheme and employing the microscopic system parameters, whereas
for $\omega_\text{mf}$ we instead inserted the effective macroscopic rates using our Monte Carlo 
algorithm, without implementing the time rescaling of Eq.~(\ref{eqs:timeScaling}). 
The data shown in this plot indicate that solving the mean-field equations with the macroscopic rates yields 
population oscillation frequencies that are considerably closer to the actual frequencies measured in the 
stochastic lattice simulations. 
However, some residual discrepancies remain, highlighting that fluctuation corrections to the characteristic 
oscillation frequency extend beyond mere parameter renormalizations \cite{intro12}.

\section{\label{section:Cyclic results} Three-species cyclic dominance}

In this section we discuss cyclic predation models, applied prominently in evolutionary game theory, where 
we restrict ourselves to three competing species $A_1$, $A_2$, and $A_3$ \cite{mf7, intro10, RPS1, 
RPS2, RPS3, RPS4, RPS5, RPS6, RPS7, RPS8, ML1, ML2, ML3, ML4, ML5WithRPSLimit, ML6, ML7, ML8, 
cyclic1, cyclic2, cyclic3, cyclic4, cyclic5, cyclic6, cyclic7, HongPaper}.
Individuals from species $A_1$ predate on $A_2$, $A_2$ in turn predate on $A_3$, and $A_3$ on 
$A_1$, completing the dominance cycle. 
Specifically, we study and compare two popular variants: 
In the ``rock-paper-scissors" (RPS) model, the nonlinear predation reaction is represented using the usual 
Lotka--Volterra prey-to-predator replacement, and hence the total particle number remains strictly 
conserved.
As a consequence, spatially extended RPS systems are prohibited to spontaneously generate 
spatio-temporal patterns, and merely display species clustering \cite{ML5WithRPSLimit, HongPaper, ML3,
ML7, RPS5}.
The second model version separates the predation and reproduction processes, thus lifting the total
population conservation law.
In such May--Leonard models, provided their spatial extension is sufficiently big, one observes the 
emergence of spiral patterns in a parameter regime where solutions with uniform particle distributions
become unstable against small inhomogeneous perturbations. 
In the following, we apply our technique for measuring the macroscopic reaction rates to both cyclic 
dominance model variations. 
In this paper, we address only symmetric cyclic model realizations, in which the various rates for all three 
species are chosen the same; therefore, the system features a discrete $\mathbb{Z}_3$ permutation 
symmetry.

\subsection{Rock-paper-scissors (RPS) / cyclic Lotka--Volterra model}

\begin{figure}[t]
    	\includegraphics[width=\columnwidth]{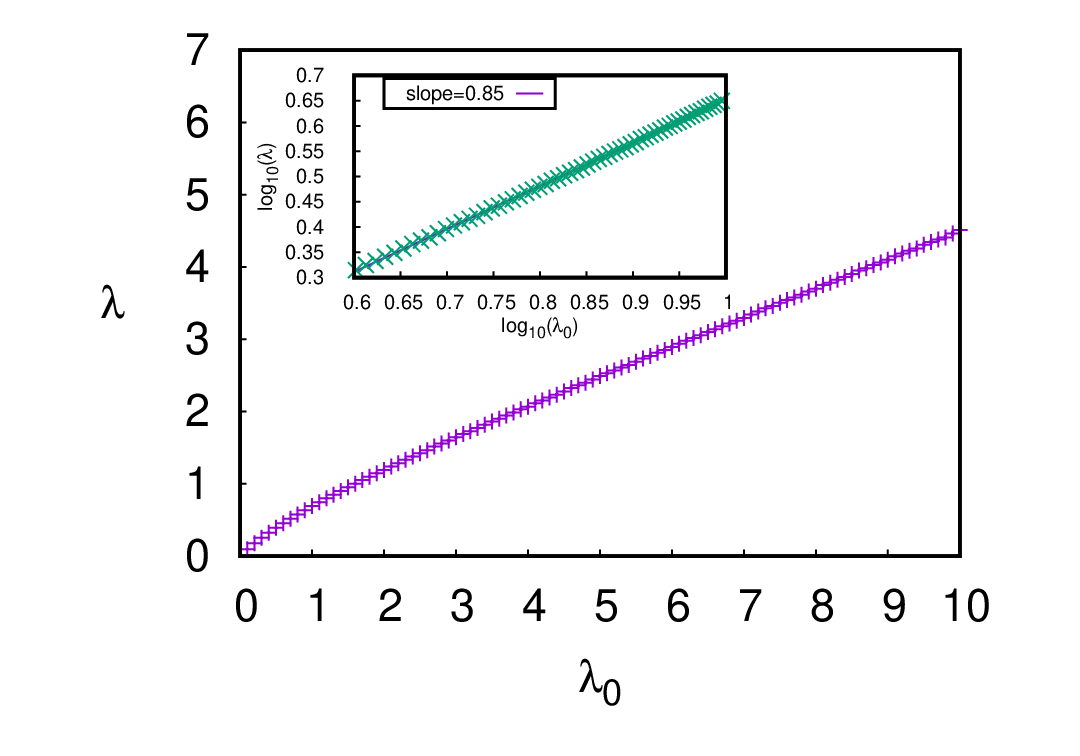} 
\caption{Macroscopic predation coupling $\lambda$ vs. its microscopic counterpart $\lambda_0$ in the
	stochastic rock-paper-scissors / cyclic Lotka--Volterra model on a two-dimensional square lattice (with 	
	periodic boundary conditions) of length $L = 150$, with equal initial densities $a_i(0) = 1/3$.
	The inset displays the same data in double-logarithmic form, demonstrating an algebraic functional
	dependence with power law exponent $\kappa \approx 0.85$. 
	The data were averaged over $100$ independent runs.}
\label{fig:RPS}
\end{figure}
\begin{figure*}[t]
    	\includegraphics[width=0.95\columnwidth]{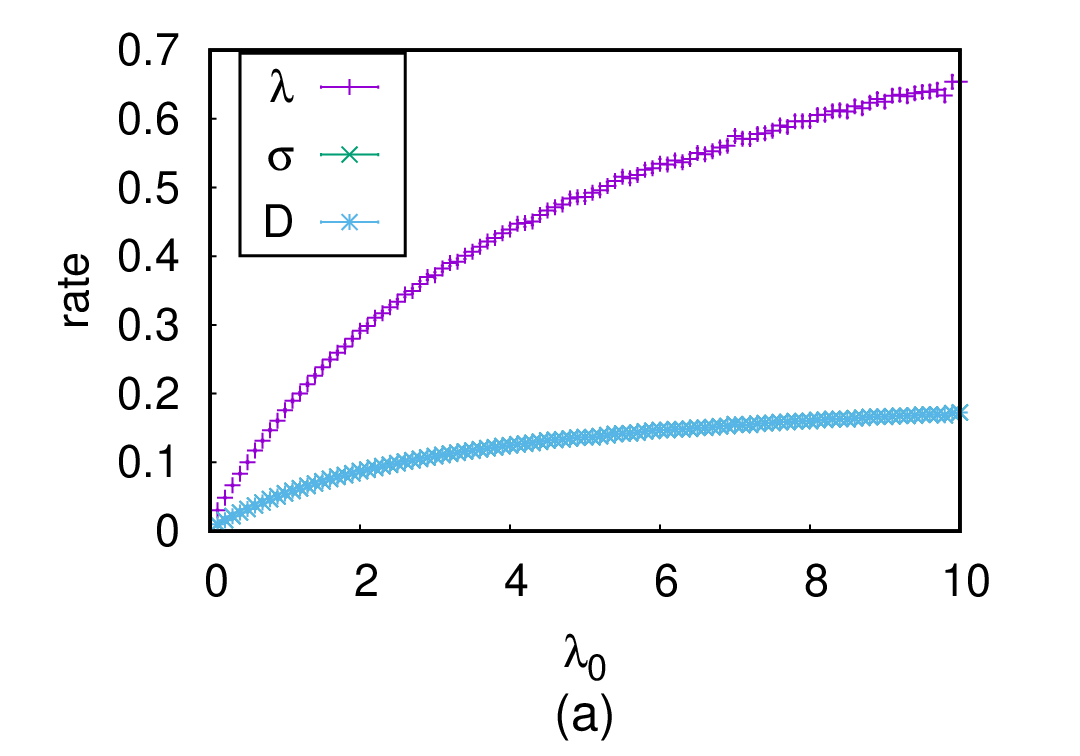} \
    	\includegraphics[width=0.95\columnwidth]{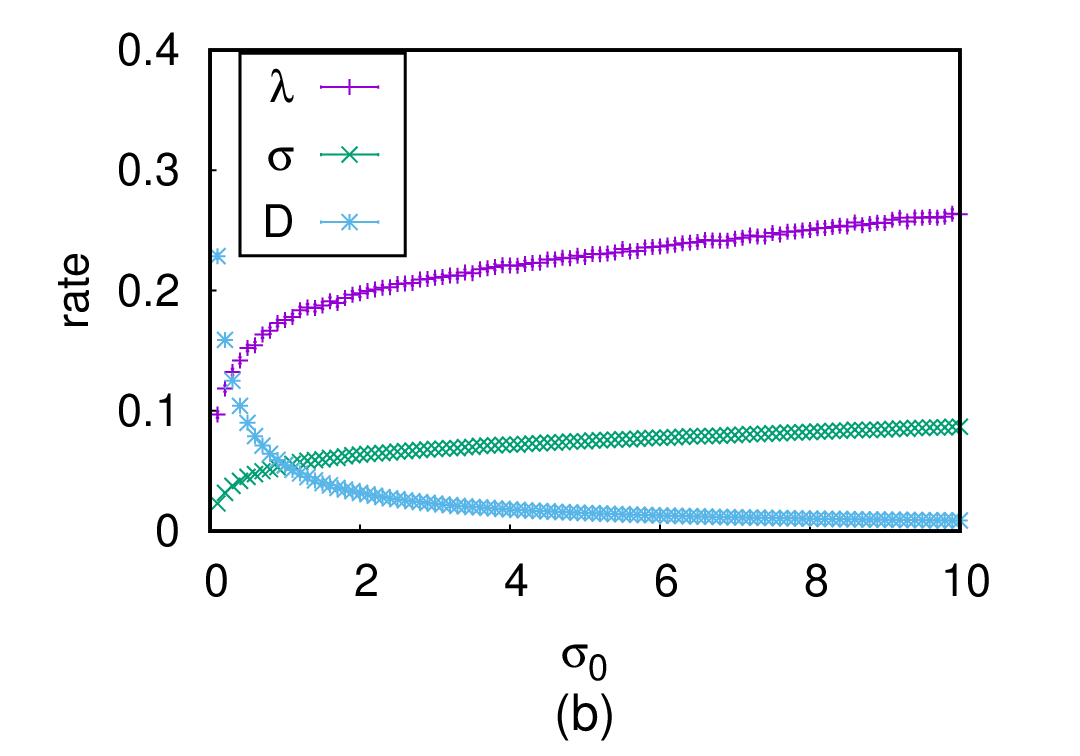}  
\caption{Macroscopic predation coupling $\lambda$, birth rate $\sigma$, and diffusivity $D$ for the
	stochastic May--Leonard model on a two-dimensional square lattice (with periodic boundary 
	conditions) of length $L = 150$ and equal initial densities $a_i(0) = 0.3$ as functions of the 
	microscopic parameters (a) $\lambda_0$, with fixed $\sigma_0 = 1$; (b) $\sigma_0$, with fixed 
	$\lambda_0 = 1$. 
	The data were averaged over $20$ independent simulation runs.}
\label{fig:ML}
\end{figure*}

The RPS or cyclic Lotka--Volterra model consists solely of the predation processes 
$A_i +A_{i+1} \xrightarrow{\lambda} 2 \, A_i$, where $i \in {1,2,3}$ and the index $i$ is cyclic, i.e., 
$i = 4$ is to be identified with $i = 1$. 
This model is implemented as a set of stochastic processes on a lattice, as before; however, no on-site 
restrictions are needed due to the conservation law. 
We implement particle transport via nearest-neighbor exchange reactions, and initialize the lattice with 
each site holding precisely one particle. 
Since no on-site restrictions apply, the exchange processes are linear, and therefore remain unaltered by
fluctuations and correlations; we set the microscopic exchange rate to $D_0 = 1$.
Thus we need not investigate the dependence of the effective particle exchange rates on the microscopic 
parameters, but only explore how the macroscopic reactive coupling $\lambda$ depends on its 
microscopic counterpart $\lambda_0$. 
The prescribed $\mathbb{Z}_3$ symmetry is of course preserved under coarse-graining; consequently
all three species will be governed by identical macroscopic parameters (this assertion was explicitly 
checked in our simulations).
The measured macroscopic coupling constant is multiplied by the time rescaling factor 
$(1 + \lambda_0)$, which ensures that $D = 1$ regardless of the value of $\lambda_0$.

In Fig.~\ref{fig:RPS}, the resulting macroscopic predation coupling $\lambda$ is plotted against its 
microscopic counterpart, the propensity $\lambda_0$. 
We observe that always $\lambda < \lambda_0$, indicating that spatial correlations reduce the effective
predation rate.
Each of the three competing species organize in inert clusters, as in two-species pair annihilation 
(Sec.~\ref{section:AnnihilationResults}), and predation reactions can only occur at their interfaces.
It turns out that the function $\lambda(\lambda_0) \sim \lambda_0^\kappa$ can be fitted to a power law 
(ignoring the first forty data points), as shown in the figure inset, with exponent $\kappa \approx 0.85$.

\subsection{May--Leonard model}

The May--Leonard model variant for three-species cyclic competition breaks the RPS local conservation law 
for the total population number by splitting the predation and reproduction reactions into independent 
stochastic processes: $A_i + A_{i+1} \xrightarrow{\lambda} A_i$ and $A_i \xrightarrow{\sigma} 2 A_i$.
We again implement a single-particle per-site constraint to control the birth reactions with rate 
$\sigma_0$. 
Since empty sites can be created by the predation reaction, explicit diffusion is incorporated through
nearest-neighbor hopping processes rather than particle exchange reactions. 
Yet hopping transport is of course limited by the on-site single-particle occupation restriction. 
Our symmetric spatially extended stochastic May--Leonard model thus comprises three coarse-grained 
parameters, $\sigma$, $\lambda$, and $D$. 
As in the RPS model, we set the microscopic hopping propensity $D_0 = 1$, and all macroscopic 
parameters are rescaled by the temporal adjustment factor $(1 + \lambda_0 + \sigma_0)$. 

Figure~\ref{fig:ML} shows the dependence of the macroscopic nonlinear coupling $\lambda$, birth rate
$\sigma$, and diffusivity $D$ on the microscopic parameters $\lambda_0$ (a) and $\sigma_0$ (b). 
Since both hopping and birth processes are conditioned on the chosen neighboring site to be empty, 
their macroscopic rate ratio $\sigma / D$ remains constant, precisely as for the prey population in the 
Lotka--Volterra system of Sec.~\ref{section:PredatorPreyResults}.
As we set $\sigma_0 = D_0 = 1$ in Fig.~\ref{fig:ML}(a), the graphs for $\sigma(\lambda_0)$ and 
$D(\lambda_0)$ coincide.
One observes nontrivial functional dependences of the macroscopic rates on the microscopic parameters:
Raising the predation propensity $\lambda_0$ increases the density of empty sites, which in turn enhances
$\sigma$ and $D$, as seen in Fig.~\ref{fig:ML}(a). 
Similarly, larger values of $\sigma_0$ tend to generate more predator-prey neighbor pairs, which is why 
the coarse-grained predation coupling $\lambda$ increases as a function of $\sigma_0$. 
In contrast, increasing $\sigma_0$ causes the lattice to saturate and therefore diminishes the density of
vacancies and consequently the effective diffusivity $D$. 
As for the RPS model, $\lambda(\lambda_0) < \lambda_0$ and likewise 
$\sigma(\sigma_0) < \sigma_0$; in fact both renormalized macroscopic rates are much smaller than 
their microscopic counterparts in the May--Leonard system.

\begin{table}[t]
\centering
\begin{tabular}{|c|c|c|}
  	\hline $\kappa$ & $\sigma_0$ & $\lambda_0$ \\
	\hline $\sigma$ & 0.17 & 0.33 \\
  	\hline $\lambda$ & 0.17 & 0.41 \\
  	\hline $D$ & -0.83 & 0.33 \\
  	\hline
\end{tabular}
\caption{Approximate scaling exponents obtained by power law fits (ignoring the first forty data points)
	of the macroscopic parameters $\sigma$, $\lambda$, and $D$ (matrix rows) for the stochastic 
	May--Leonard model on a square lattice as functions of the microscopic predation rate $\lambda_0$ 
	and birth rate $\sigma_0$ (columns).}
\label{tab:exponents_table}
\end{table}
Akin to the RPS model, the macroscopic rates $\sigma$, $\lambda$, and $D$ for the May--Leonard 
system as functions of the microscopic input parameters $\sigma_0$ and $\lambda_0$ can be 
approximately fitted to power laws . 
The values of the six resulting effective scaling exponents $\kappa$ are listed in 
Table~\ref{tab:exponents_table}. 
We note that the exponent for the function $\lambda(\lambda_0)$ comes out much smaller for the
May--Leonard as compared to the RPS model, suggesting that the emerging spiral patterns protect each 
species against predation more effectively than clustering. 
Since their ratio is preserved under coarse-graining, both $\sigma$ and $D$ exhibit identical power laws 
as functions of $\lambda_0$. 
Surprisingly, our results indicate that also $\lambda$ and $\sigma$ show the same dependence on 
$\sigma_0$. 
To explain this feature, we recall that the macroscopic rate equations for the symmetric May--Leonard 
model yield the stationary densities $a_s = \sigma / \lambda$.
However, as $\sigma_0$ increases past a certain point, the stationary densities $a_s$ must saturate at 
the value $1/3$, owing to the prescribed on-site restrictions. 
As a consequence, once $a_s \approx 1/3$, we find that $\lambda = 3 \, \sigma$ and both macroscopic
parameters obey the same scaling, and indeed, their ratio $3$ is confirmed in Fig.~\ref{fig:ML}(b).
Finally, since the (anti-)correlations induced by the on-site restrictions preserve the ratio 
$D(\sigma_0) / \sigma(\sigma_0) = D_0 / \sigma_0 = 1 / \sigma_0$, the associated power law 
exponents should differ by $1$, as is indeed borne out nicely by the data in 
Table~\ref{tab:exponents_table}.
It should be noted that these scaling exponents are not universal, as we found that simulating stochastic
May--Leonard models with different microscopic rates (and spatial dimensions) produces different power 
laws.  
However, we checked that the exponents $\kappa$ do not depend on the system size (provided it is
sufficiently large).

\section{\label{section:conclusions} Summary and conclusions}

Reaction-diffusion systems are often approximately represented through mean-field ordinary or partial 
differential equations that largely ignore the effects of stochastic fluctuations and spatial and temporal
correlations. 
While certain systems may qualitatively still be modeled by means of such coupled rate equations, e.g.,
in sufficiently high dimensions or on strongly connected networks, the effective coarse-grained continuum 
rates appearing in these nonlinear dynamical equations are hardly ever given by the density- or 
scale-independent microscopic system parameters. 
In this article, we have proposed a computational technique that employs agent-based stochastic Monte 
Carlo lattice simulations to extract the effective renormalized macroscopic reaction rates. 
The results numerically obtained by this method were first tested for simple restricted birth processes, 
subject to on-site restrictions on the lattice.
We saw that the system is initially described by the growth equation with the microscopic rate value, 
whereas (only) the long-time asymptotic is described by a logistic equation with a modified effective birth 
rate.
We then compared our simulation data against the well-known results for diffusion-limited single-species 
binary coagulation and two-species pair annihilation models, thus validating the effects of particle 
(anti-)correlations on the long-time scaling properties of the macroscopic annihilation rates, as a function 
either of time or the diminishing particle densities. 
Furthermore, the usefulness of computing the macroscopic rates was highlighted by providing a 
straightforward metric tool for whether the system has reached its asymptotic scaling regime. 
This was accomplished by probing the temporal scaling of the macroscopic annihilation rate. 

Next, we applied our technique to the paradigmatic stochastic Lotka--Volterra model for predator-prey 
competition and coexistence on a square lattice, for which we measured various effective rates for the 
involved stochastic processes.
We found that representing the system through the coupled mean-field rate equations with the measured 
macroscopic parameters instead of the original microscopic ones, leads to a much improved agreement 
with the lattice simulation data. 
Yet we also detected clear renormalization effects caused by fluctuations in the system that are not 
captured by mere rate or coupling constant renormalizations.
In cyclic dominance models of three competing species, our numerical method was used to compare the 
effects of correlations on the macroscopic rate values in the cyclic Lotka--Volterra (RPS) and May--Leonard 
models. 
In both systems, one may fit the emergent coarse-grained effective rates to algebraic power-law 
dependences on the microscopic rate parameters. 
In the May--Leonard model, the effective macroscopic rate exhibits a sharper decrease as a function of its
microscopic counterpart as compared to the RPS variant, which we attribute to the presence of spiral 
spatio-temporal patterns that stabilize each species against predation events.
Finally, we were able to derive relationships between several rate scaling exponents.

The present work demonstrates the possibility of utilizing agent-based Monte Carlo simulations to compute 
effective macroscopic rates in stochastic reactive many-particle models, which in some situations can 
markedly improve fits to rate equation solutions.
Hence the technique proposed in this article may serve as a refined alternative to fitting data to rate 
equations with constant parameters, as our method allows the implementation of coarse-grained 
scale-dependent rates that incorporate the time or density dependence due to emerging correlations in the 
involved stochastic reaction processes.
In addition, one may directly probe the large-scale features emerging in complex interacting many-particle
systems owing to the interplay of stochastic reaction processes with time-dependent intrinsic constraints.

\begin{acknowledgments}
We would like to thank Matthew Asker, Kenneth Distefano, Llu\'is Hernandez-Navarro, Mauro Mobilia, 
Michel Pleimling, Alastair Rucklidge, Louie Hong Yao, and Canon Zeidan for fruitful discussions regarding 
this work.
This research was supported by the U.S. National Science Foundation, Division of Mathematical Sciences 
under Award No. NSF DMS-2128587.
\end{acknowledgments}

\bibliographystyle{apsrev4-2}
\bibliography{mybib}

\end{document}